\documentclass[a4paper,journal,romanappendices]{IEEEtran}
\usepackage{amssymb}
\usepackage{amsmath}
\usepackage{mathrsfs}
\usepackage{rotating}
\usepackage{overpic}
\usepackage{algorithm}
\usepackage{algorithmic}
\usepackage{multirow}
\usepackage{array}
\usepackage{graphicx}
\usepackage[caption=false]{subfig}
\usepackage{bm}
\usepackage{color}
\usepackage{tikz}
\usepackage{epstopdf}
\usepackage{stfloats}
\usepackage{cite}
\usepackage{tikz}
\usepackage{amsmath}
\usepackage{psfrag}
\usepackage{acronym}
\usepackage{enumerate}
\usetikzlibrary{arrows}
\usetikzlibrary{decorations}

\definecolor{BLUE}{rgb}{0,0,1}

\newtheorem{proposition}{Proposition}

\newtheorem{definition}{Definition}

\newcommand{\tr}[1]{{\rm Tr}\left\{#1\right\}}

\newcommand{\diag}[1]{{\rm diag}\left\{#1\right\}}

\acrodef{qem}[QEM]{quantum error mitigation}
\acrodef{qpr}[QPR]{quasi-probability representation}
\acrodef{kkt}[KKT]{Karush-Kuhn-Tucker}
\acrodef{nisq}[NISQ]{noisy intermediate-scale quantum}
\acrodef{fir}[FIR]{finite impulse response}
\acrodef{qsa}[QSA]{quantum search algorithm}
\acrodef{qaoa}[QAOA]{quantum approximate optimization algorithm}
\acrodef{vd}[VD]{virtual distillation}
\acrodef{mimo}[MIMO]{multiple-input mutliple-output}

\acrodefplural{qsa}[QSA]{quantum search algorithms}

\usepackage{winsnotation}
\usepackage[braket]{qcircuit}
\usepackage{mathrsfs}
\usepackage{makecell}
\usepackage{rotating}
\newcommand{\Sop}[1]{{\mathcal{#1}}}

\definecolor{myblue}{rgb}{0.2,0.2,0.7}

\begin{document}
\title{Quantum Error Mitigation Relying on \\ Permutation Filtering}
\author{Yifeng Xiong, Soon Xin Ng, \IEEEmembership{Senior Member, IEEE}, and Lajos Hanzo, \IEEEmembership{Fellow, IEEE}
\thanks{Authors are with School of Electronics and Computer Science, University of Southampton, SO17 1BJ, Southampton (UK). The insightful comments of Balint Koczor are gratefully acknowledged by
the authors.}
\thanks{L. Hanzo would like to acknowledge the financial support of the Engineering and Physical Sciences Research Council projects EP/N004558/1, EP/P034284/1, EP/P034284/1, EP/P003990/1 (COALESCE), of the Royal Society's Global Challenges Research Fund Grant as well as of the European Research Council's Advanced Fellow Grant QuantCom. This work is also supported in part by China Scholarship Council (CSC).}
}
\maketitle
\begin{abstract}
Quantum error mitigation (QEM) is a class of promising techniques capable of reducing the computational error of variational quantum algorithms tailored for current noisy intermediate-scale quantum computers. The recently proposed permutation-based methods are practically attractive, since they do not rely on any \textit{a priori} information concerning the quantum channels. In this treatise, we propose a general framework termed as permutation filters, which includes the existing permutation-based methods as special cases. In particular, we show that the proposed filter design algorithm always converge to the global optimum, and that the optimal filters can provide substantial improvements over the existing permutation-based methods in the presence of narrowband quantum noise, corresponding to large-depth, high-error-rate quantum circuits.
\end{abstract}


\section*{Notations}
\begin{itemize}
\item Scalars, vectors and matrices are represented by $x$, $\V{x}$, and $\M{X}$, respectively. Sets and operators are denoted as $\Set{X}$ and $\Sop{X}$, respectively.
\item The notations $\V{1}_n$, $\V{0}_{n}$, $\V{0}_{m\times n}$, and $\M{I}_{k}$, represent the $n$-dimensional all-one vector, the $n$-dimensional all-zero vector, the $m\times n$ dimensional all-zero matrix, and the $k\times k$ identity matrix, respectively.
\item The notation $\|\V{x}\|_p$ represents the $\ell_p$-norm of vector $\V{x}$, and the subscript may be omitted when $p=2$. For matrices, $\|\M{A}\|_p$ denotes the matrix norm induced by the corresponding $\ell_p$ vector norm.
\item The notation $[\M{A}]_{i,j}$ denotes the $(i,j)$-th entry of matrix $\V{A}$. For a vector $\V{x}$, $[\V{x}]_i$ denotes its $i$-th element. The submatrix obtained by extracting the $i_1$-th to $i_2$-th rows and the $j_1$-th to $j_2$-th columns from $\M{A}$ is denoted as $[\M{A}]_{i_1:i_2,j_1:j_2}$. The notation $[\M{A}]_{:,i}$ represents the $i$-th column of $\M{A}$, and $[\M{A}]_{i,:}$ denotes the $i$-th row, respectively.
\item The trace of matrix $\M{A}$ is denoted as ${\mathrm{Tr}}\{\M{A}\}$.
\item The notation $\M{A}\otimes \M{B}$ represents the Kronecker product between matrices $\M{A}$ and $\M{B}$.
\item Pure states are denoted by ``kets'' $\ket{\psi}$, and their dual vectors are denoted by ``bras'' $\bra{\psi}$.
\end{itemize}

\section{Introduction}
\IEEEPARstart{Q}{uantum} technologies have entered the era of \ac{nisq} computation \cite{nisq}. These computers typically rely on dozens to a few hundreds of qubits. Remarkably, \ac{nisq} computers based on both superconductive \cite{quantum_supremacy} and photonic technologies \cite{quantum_supremacy_photon}, have shown quantum advantage in computing certain tasks.

However, \ac{nisq} computers may not afford fully fault-tolerant operations \cite{threshold_thm} enabled by quantum error correction codes \cite{qecc,qit,qecc_survey,qtecc,extra_ecc}, since the qubit overhead is still prohibitive for state-of-the-art devices. Consequently, quantum algorithms requiring long coherence time, such as the quantum phase estimation algorithm \cite{ncbook} and the quantum amplitude amplification \cite{qaa1,qaa2}, may not be practical for quantum computers available at the time of writing. Notably, these algorithms are often used as subroutines of more sophisticated quantum algorithms relying on the assumption of fault-tolerance, including Shor's factoring algorithm \cite{shor} and Grover's search algorithm \cite{grover,extra_qsa,extra_qsa2}. This suggests that a paradigm shift both for algorithm design and for error control techniques might be necessary for \ac{nisq} computers.

\begin{figure}[t]
\centering
\begin{overpic}[width=0.495\textwidth]{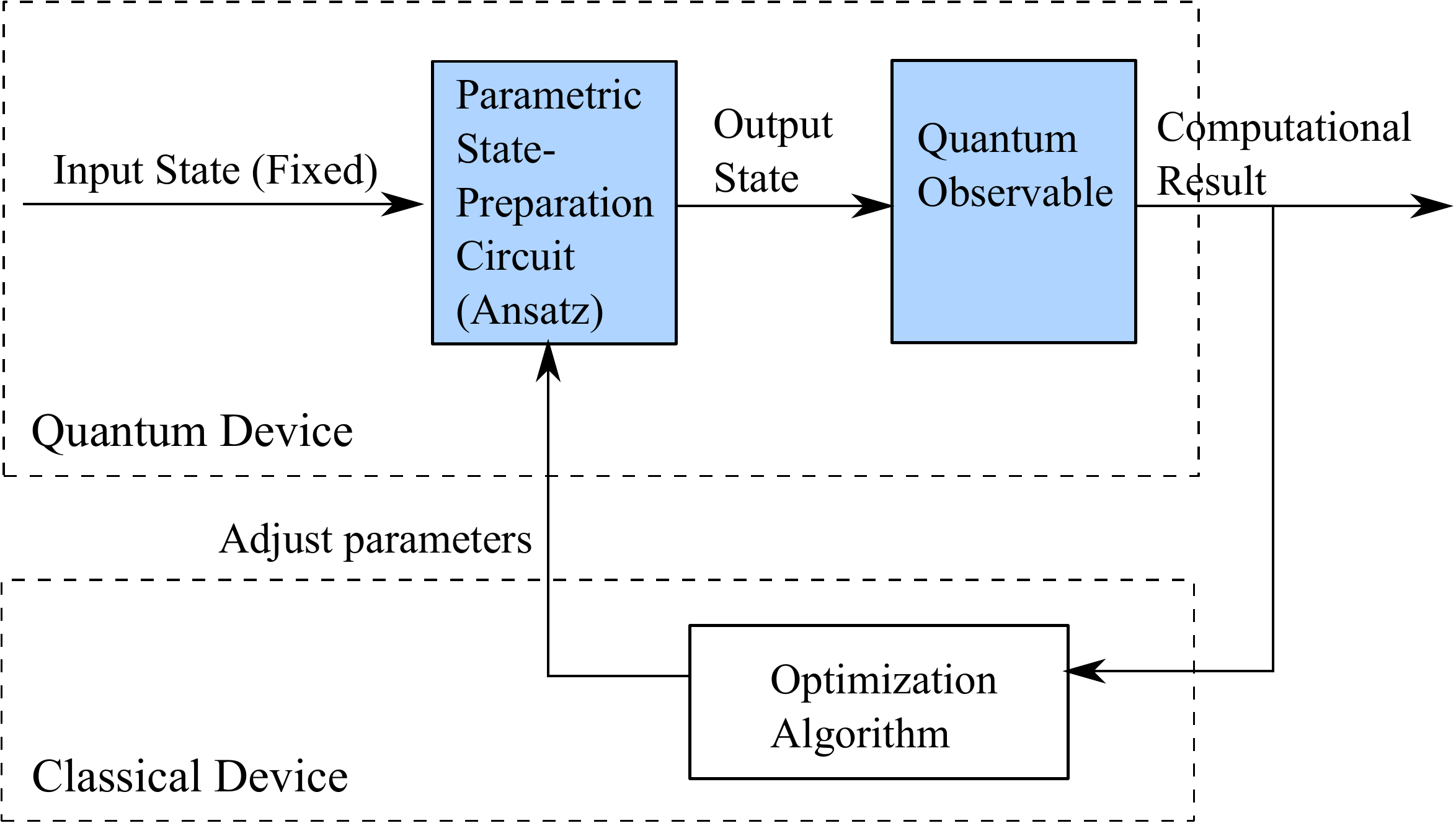}
\put(68,38){\color{myblue}\scriptsize $\Sop{H}$}
\put(39,18.5){\color{myblue}\scriptsize $\V{\theta}^{(l+1)}=\nu\left[J(\V{\theta}^{(l)}),\V{\theta}^{(l)}\right]$}
\put(47,38){\color{myblue}\scriptsize $\ket{\psi(\V{\theta})^{(l)}}$}
\put(88,37){\color{myblue}\scriptsize $J(\V{\theta}^{(l)})$}
\put(7,38){\color{myblue}\scriptsize e.g., $\ket{0}^{\otimes N_{\rm q}}$}
\end{overpic}
\caption{The workflow of a typical variational quantum algorithm.}
\label{fig:vqa}
\end{figure}
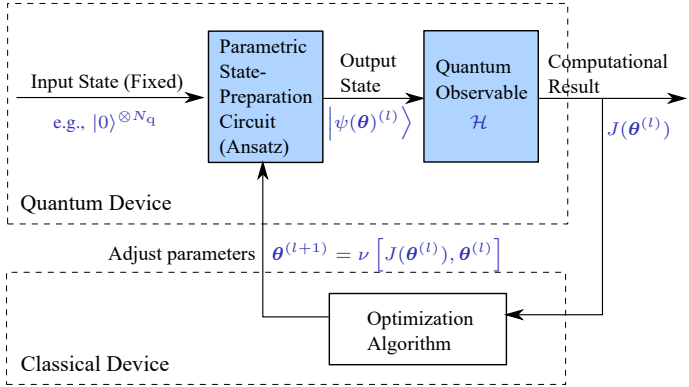

As proposed in \cite{vqe}, variational quantum algorithms \cite{vqe,vqe2,qaoa,performance_qaoa,vqlinear} constitute one of the new algorithm design paradigms harnessing the computational power of \ac{nisq} computers without relying on quantum error correction techniques, including the celebrated variational quantum eigensolver \cite{vqe} and the \ac{qaoa} \cite{qaoa}. Specifically, the eigenvalue evaluation subroutine, which is typically realized using the quantum phase estimation algorithm in ``traditional'' quantum algorithms, is implemented in variational quantum algorithms by directly measuring the corresponding quantum observables \cite{vqe_theory}. The workflow of a typical variational quantum algorithm is portrayed in Fig.~\ref{fig:vqa}. To elaborate further, these algorithms aim for designing parametric state preparation circuits using an iterative, hybrid quantum-classical optimization procedure that outputs (approximate) the eigenstates of the specific Hamiltonian encoding the computational task. The eigenvalues can then be estimated by directly measuring the observables. By contrast, in the quantum phase algorithm, the Hamiltonian simulation \cite{hamiltonian_simulation} subroutine is executed on the order of $O(1/\epsilon)$ times, where $\epsilon$ denotes the required accuracy, hence the coherence time requirements of physical qubits are more strict than those of variational algorithms.

Despite the fact that the parametric state preparation circuits in variational quantum algorithms have relatively short depth (compared to that of the quantum phase estimation algorithm), they can still be so deep that the imperfections of the circuits accumulate to an amount that lead to significant computational errors. This calls for effective error control methods that do not rely on the fault-tolerant scheme requiring many qubits. One of the most promising error control strategies conceived for \ac{nisq} computers is \ac{qem} \cite{qem} tailored for variational quantum algorithms. Typically, \ac{qem} methods mitigate the error with the aid of classical post-processing. This reduces both the additional errors introduced by error control quantum operations as well as the qubit overhead represented by the number of ancillas used in quantum error correction.

\begin{table*}[t]
\centering
\footnotesize
\caption{Comparisons between different \ac{qem} methods.}
\footnotesize
\label{tbl:compare_qem}
\footnotesize
\begin{tabular}{|l|l|l|l|}
\hline
 & \textbf{Main overhead} & \textbf{Prior knowledge required} & \textbf{Remark}\\
\hline
\makecell[l]{Zero-noise Extrapolation  \cite{qem,practical_qem,extrapolation2}} & Sampling overhead & No  & Requires pulse-level control \\
\hline
\makecell[l]{Channel inversion  \cite{qem,qem_exp,qem_suguru}} & Sampling overhead & \makecell[l]{Channel estimation \\ (gate set tomography)}  & \makecell[l]{Has error floor due to \\ imperfect channel estimation}\\
\hline
Learning-based  \cite{clifford_regression,data_driven_qem} & Sampling overhead & \makecell[l]{Pre-training on \\ certain circuit sets}  &\\
\hline
\makecell[l]{Symmetry verification \cite{nisq_error2,symmetry2}} & \makecell[l]{Sampling overhead, \\ qubit overhead}& \makecell[l]{Type of symmetries \\ in the computational task}  & \textbf{Symmetry-based}\\
\hline
\makecell[l]{Virtual distillation  \cite{koczor,permutation2}} & \makecell[l]{Sampling overhead, \\ qubit overhead}& No  & \textbf{Symmetry-based}\\
\hline
\textbf{This treatise} & \makecell[l]{Sampling overhead, \\ qubit overhead}& No  & \makecell[l]{\textbf{Improves the accuracy of VD} \\ \textbf{at a similar overhead}}\\
\hline
\end{tabular}
\end{table*}

Broadly speaking, there have been four types of \ac{qem} methods. One of them collects the computational results produced by circuits having different error rates, and then extrapolates the results to the point where the error rate tends to zero \cite{qem,practical_qem,extrapolation2}. Another idea is to construct a set of probabilistic quantum circuits effectively implementing the inverse of the error operator (also known as the quantum channel) \cite{qem,qem_exp,qem_suguru}. There have also been learning-based methods that mitigate the error of practical sophisticated circuits using statistical models that pre-trained on Clifford circuits, which have known efficient simulation algorithms on classical computers \cite{clifford_regression,data_driven_qem}. The fourth concept exploits the symmetry (redundancy) of the quantum states or the computational task itself for mitigating the error rate, by preventing the states that do not satisfy certain symmetry conditions from contributing to the computational result \cite{nisq_error2,symmetry2}. The characteristics of the \ac{qem} methods are summarized in Table \ref{tbl:compare_qem}. In general, these methods are not mutually exclusive in practical applications. Instead, potentially beneficial combinations have been conceived in \cite{koczor}. For a comprehensive comparison between these methods, interested readers may refer to \cite{unifying}.

\begin{figure*}[t]
\centering
\begin{overpic}[width=0.7\textwidth]{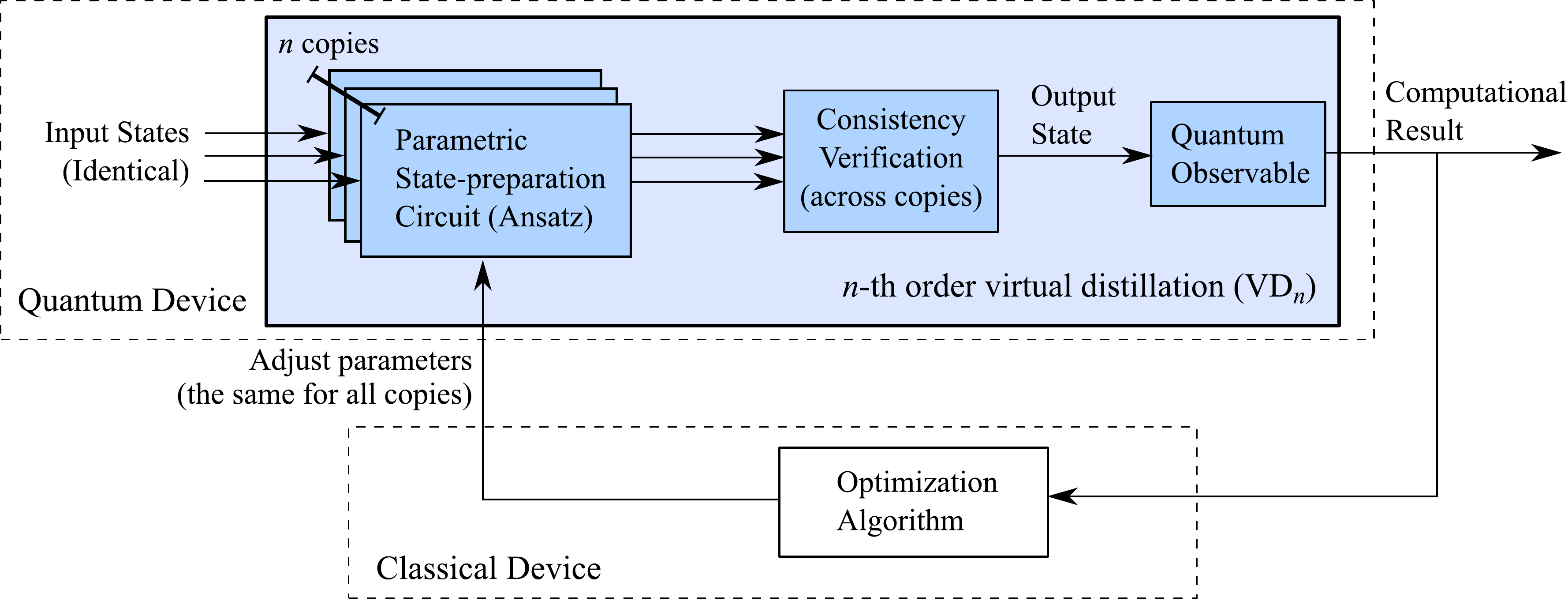}
\end{overpic}
\caption{An $n$-th order virtual distillation method (relying on $n$ copies of the parametric state-preparation circuits) applied to a variational quantum algorithm.}
\label{fig:vqa_vd}
\end{figure*}

Recently, a new class of symmetry-aided \ac{qem} methods, namely the \ac{vd} \cite{koczor,permutation2}, has been proposed, which relies on the permutation symmetry of quantum states. To elaborate, they prepare multiple copies of the same quantum state, and filter out the components in the states that are not identical across all copies, as shown in Fig.~\ref{fig:vqa_vd}. The observables are then measured on one of the copies. Compared to previous \ac{qem} methods, the advantage of these techniques is that they do not require \textit{a priori} knowledge about the quantum channels, and that the symmetry of the states can be easily manipulated by adjusting the number of copies.

\begin{figure*}[t]
\centering
\begin{overpic}[width=0.85\textwidth]{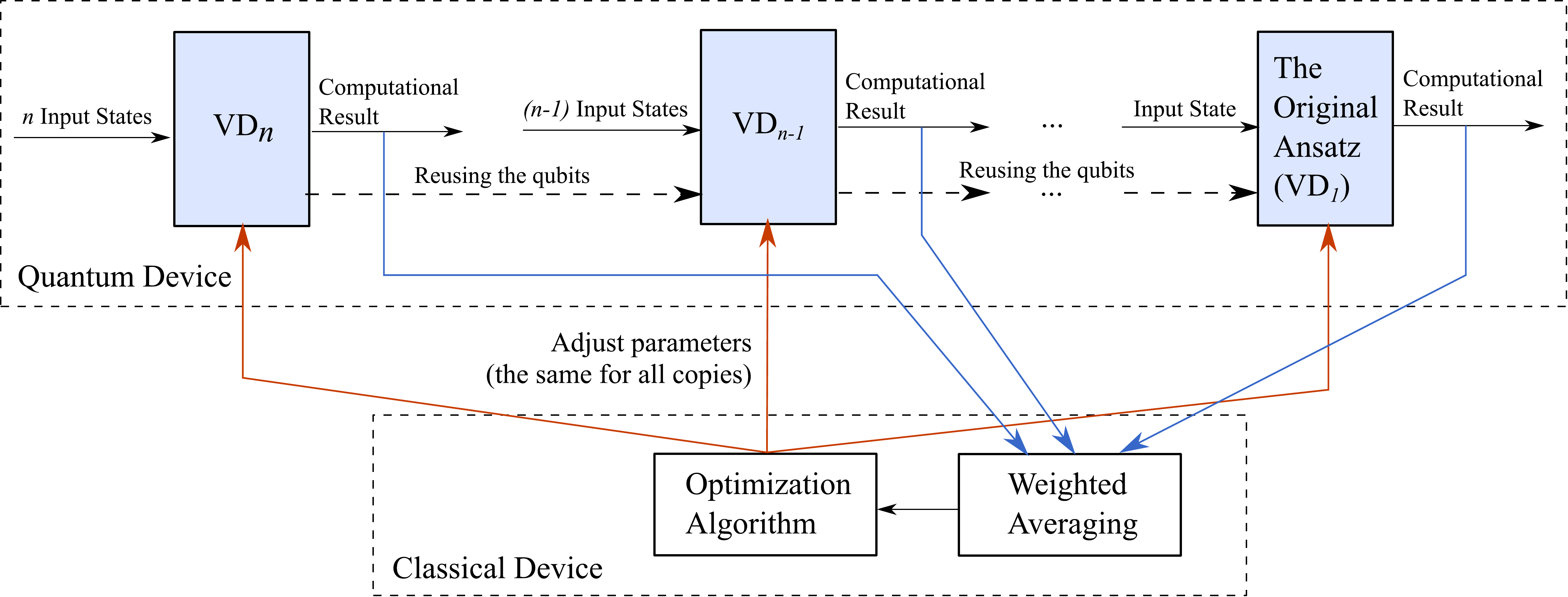}
\end{overpic}
\caption{An $n$-th order permutation filter proposed in this treatise applied to a variational quantum algorithm.}
\label{fig:vqa_pfilter}
\end{figure*}

From the spectral analysis perspective of quantum states, when the noise is not extremely strong, the dominant eigenvector of the output state serves as a good approximation of the ideal noise-free output state \cite{koczor}. In this sense, the permutation-based \ac{qem} methods may be viewed as high-pass filters in the spectral domain. In this treatise, we generalize this idea by proposing a general framework for designing optimal filters in the spectral domain of quantum states. These filters assume a similar form as the \ac{fir} filters widely used in classical signal processing tasks, by computing a weighted average over the outputs of multiple virtual distillation circuits of different orders, as shown in Fig.~\ref{fig:vqa_pfilter}. Our novel contributions are summarized below.

\begin{itemize}
  \item We propose a general permutation filter design framework, including the functional form of the filters and the performance metric to be optimized. We will show that existing permutation-based \ac{qem} methods may be viewed as specific cases of permutation filters.
  \item We propose an algorithm for optimal permutation filter design. In particular, we show that the local optimum of the optimization problem is unique, hence the global optimal solution is attainable by the proposed algorithm.
  \item We show that permutation filters are particularly efficient in combating narrowband noise. Specifically, they are capable of providing an error-reduction improvement scaling polynomially with respect to the noise bandwidth, compared to the existing permutation-based \ac{qem} methods.
  \item We also show that the noise bandwidth decreases exponentially with the depth of the quantum circuit. This suggests that the proposed permutation filters can be used for supporting the employment of quantum circuits having an increased depth without degrading their fidelity.
\end{itemize}

The rest of this treatise is organized as follows. In Section~\ref{sec:preliminaries} we provide a brief introduction to variational quantum algorithms and permutation-based \ac{qem} methods. In Section~\ref{sec:filter}, we describe the permutation filter as well as its design algorithm. Then, in Section~\ref{sec:performance} we analyze the error-reduction performance of permutation filters. The results are further illustrated using numerical results in Section~\ref{sec:numerical}. Finally, we conclude the paper in Section~\ref{sec:conclusions}.

\section{Preliminaries}\label{sec:preliminaries}
\subsection{Variational Quantum Algorithms}
Variational quantum algorithms constitute a class of hybrid quantum-classical algorithms \cite{hqc_survey} tailored for \ac{nisq} computers, which aim for solving optimization problems of the following form
\begin{equation}
\begin{aligned}
\hat{\V{\theta}} &= \mathop{\arg\min}_{\V{\theta}}J(\V{\theta}), \\
J(\V{\theta})&=\bra{\psi(\V{\theta})}\Sop{H}\ket{\psi(\V{\theta})},
\end{aligned}
\end{equation}
where $\Sop{H}$ is the Hamiltonian encoding the optimization cost function, and the mapping from $\V{\theta}$ to the quantum state $\ket{\psi(\V{\theta})}$ is implemented by a parametric state preparation circuit, also known as the \textit{ansatz} \cite{ansatz}.

When we work on qubits, it is often convenient to decompose the Hamiltonian into a weighted sum of Pauli operators (so-called ``Pauli-strings'' defined in \cite{pauli_string}). In particular, a Hamiltonian acting upon $N_{\rm q}$ qubits may be expressed as

\begin{equation}\label{sum_string}
\Sop{H} = \sum_{i=1}^{4^{N_{\rm q}}} w_i \Sop{S}_i^{(N_{\rm q})},
\end{equation}
where $\Sop{S}_i^{(N_{\rm q})}$ denotes the $i$-th Pauli string acting upon $N_{\rm q}$ qubits, given by

\begin{equation}
\Sop{S}_{i}^{(N_{\rm q})}=\bigotimes_{j=1}^{N_{\rm q}} \Sop{S}_{{\rm digit}(i,j)+1}^{(1)},
\end{equation}
where ${\rm digit}(i,j)$ represents the $j$-th digit of $i$ when treated as a base-4 number. The single-qubit Pauli operators $\Sop{S}_k^{(1)},~k=1,2,3,4$, are given by
$$
\begin{aligned}
\Sop{S}_1^{(1)}=\M{S}_{\Sop{I}} &= \left[
          \begin{array}{cc}
            1 & 0 \\
            0 & 1 \\
          \end{array}
        \right],~\Sop{S}_2^{(1)}=\M{S}_{\Sop{X}} = \left[
          \begin{array}{cc}
            0 & 1 \\
            1 & 0 \\
          \end{array}
        \right], \\
\Sop{S}_3^{(1)}=\M{S}_{\Sop{Y}} &= \left[
          \begin{array}{cc}
            0 & -i \\
            i & 0 \\
          \end{array}
        \right],~\Sop{S}_4^{(1)}=\M{S}_{\Sop{Z}} = \left[
          \begin{array}{cc}
            1 & 0 \\
            0 & -1 \\
          \end{array}
        \right].
\end{aligned}
$$
The number of $t_i$ values satisfying $t_i\neq 1$ is called the weight $\omega(\Sop{S}_{p(\V{t})})$ of the Pauli string $\Sop{S}_{p(\V{t})}$, and in general we have $1<\omega(\Sop{S}_{p(\V{t})})\le N_{\rm q}$.

In variational quantum algorithms, the observation of the complicated Hamiltonian $\Sop{H}$ is implemented by a set of observations of the corresponding Pauli strings, as follows \cite{vqe_theory}:
\begin{equation}\label{decomp_observable}
\bra{\psi(\V{\theta})}\Sop{H}\ket{\psi(\V{\theta})} = \sum_{i=1}^{4^{N_{\rm q}}} w_i \bra{\psi(\V{\theta})}\Sop{S}_i^{(N_{\rm q})}\ket{\psi(\V{\theta})}.
\end{equation}
To take full advantage of the computational power of both classical and quantum devices, the variational quantum algorithms solve the optimization problem in an iterative fashion as follows (also shown in Fig.~\ref{fig:vqa}):
\begin{subequations}
\begin{align}
J(\V{\theta}^{(l)})&=\sum_{i=1}^{4^{N_{\rm q}}} w_i \bra{\psi(\V{\theta}^{(l)})}\Sop{S}_i^{(N_{\rm q})}\ket{\psi(\V{\theta}^{(l)})}, \label{cost_compute_pure}\\
\V{\theta}^{(l+1)}&=\nu\left[J(\V{\theta}^{(l)}),\V{\theta}^{(l)}\right],
\end{align}
\end{subequations}
where $\nu\left[J(\V{\theta}^{(l)}),\V{\theta}^{(l)}\right]$ is an update rule for the parameters defined by the specific algorithm. This hybrid quantum-classical optimization procedure aims for finding the optimal eigenvalue using short-depth circuits, thus avoiding the strict coherence time requirements of the quantum phase estimation algorithm.

In practice, the state preparation circuit outputs are contaminated by decoherence, which turns the output states into a mixed form. Hence, the practical version of \eqref{cost_compute_pure} is given by
\begin{equation}
\tilde{J}_{l}(\V{\theta}^{(l)})=\sum_{i=1}^{4^{N_{\rm q}}} w_i \tr{\rho(\V{\theta}^{(l)})\Sop{S}_i^{(N_{\rm q})}},
\end{equation}
where $\rho(\V{\theta}^{(l)})$ is a mixed state, as opposed to the pure state $\ket{\psi(\V{\theta})}$ of the previous discussion. Apparently, the noisy cost function $\tilde{J}_{l}(\cdot)$ would be different from the ideal cost function $J(\cdot)$, and hence their values at the specific parameter $\V{\theta}^{(l)}$ would also be different. The difference will become more significant when the state preparation circuit is more complex (i.e., either deep or involves a large number of qubits). This necessitates the employment of quantum error mitigation, which aims for ``purifying'' the mixed state $\rho(\V{\theta}^{(l)})$, in order to mitigate the contamination of the computed cost function values.

\subsection{Permutation-based Quantum Error Mitigation}
\begin{figure}[t]
\centering
\subfloat[][Swap test]{
\begin{minipage}{.42\columnwidth}
\centering
\includegraphics[width=.99\textwidth]{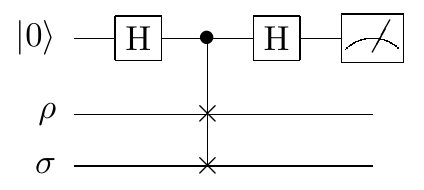}
\end{minipage}
\label{fig:swap_test}
}
\centering
\subfloat[][Permutation test]{
\begin{minipage}{.48\columnwidth}
\centering
\includegraphics[width=.99\textwidth]{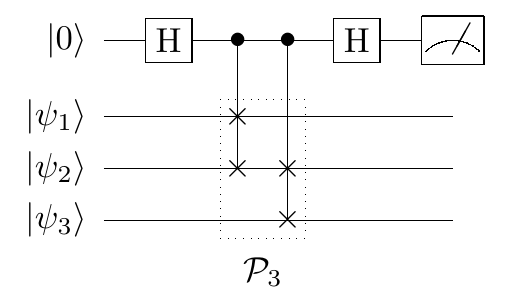}
\end{minipage}
\label{fig:perm_test}
}
\\
\subfloat[][Koczor's method]{
\begin{minipage}{.48\columnwidth}
\centering
\begin{overpic}[width=.99\textwidth]{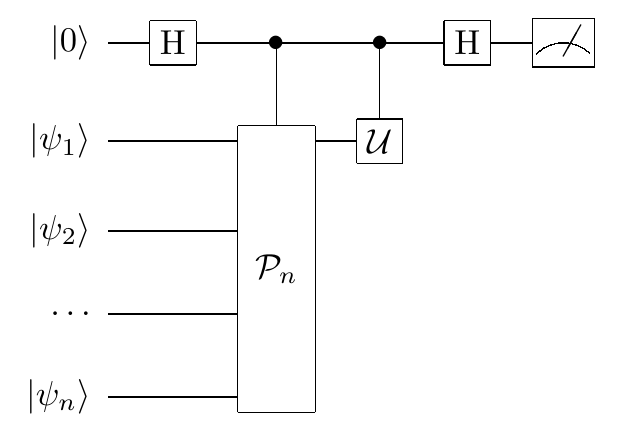}
\end{overpic}
\end{minipage}
\label{fig:koczor}
}
\subfloat[][The \ac{vd} in \cite{permutation2}]{
\begin{minipage}{.45\columnwidth}
\centering
\begin{overpic}[width=.99\textwidth]{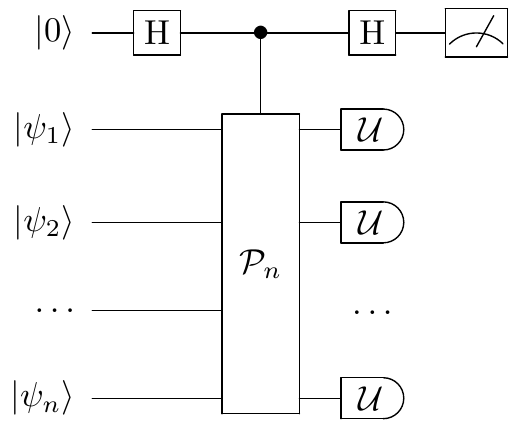}
\end{overpic}
\end{minipage}
\label{fig:vd_another}
}
\caption{Schematics of the swap test, the permutation test, and two circuit implementations of the virtual distillation method.}
    \label{fig:tests}

\end{figure}

The permutation-based quantum error mitigation philosophy is inspired by the concept of permutation tests, which constitute generalizations of the swap test \cite{swap_test}. As portrayed in Fig.~\ref{fig:swap_test}, the swap test is implemented by controlled-SWAP gates. It is widely employed for evaluating the overlap between a pair of quantum states $\rho$ and $\sigma$, since the expected value of the measurement outcome is given by $\mathrm{Tr}\{\rho\sigma\}$. Naturally, when we have two copies of the same state $\rho$, we may compute $\mathrm{Tr}\{\rho^2\}$ using the swap test.

The permutation tests, exemplified by the cyclic-shift test \cite{cyclic_shift1}, may be implemented using quantum circuits taking the form shown in Fig.~\ref{fig:perm_test}. As a generalization of the swap gate, an $n$-th order cyclic-shift circuit $\mathcal{P}_n$ taking an input of $n$ pure states $\ket{\psi_1,\psi_2,\dotsc,\psi_n}$ would output a shifted state $\ket{\psi_2,\psi_3,\dotsc,\psi_n,\psi_1}$. Note that the swap gate may be viewed as a specific case of cyclic-shift circuit, since it is equivalent to $\mathcal{P}_2$. Similar to the swap test, one may show that the expectation value of the outcome in an $n$-th order cyclic-shift test is given by $\mathrm{Tr}\{\rho^n\}$ \cite{koczor}, when the inputs are represented by $n$ copies of the same mixed state $\rho$.

Typically, when quantum circuits are contaminated by decoherence, the output state would approximately take the following form
\begin{equation}
\rho = \lambda_1 \ket{\psi}\bra{\psi} + \sum_{i=2}^{2^{N_{\rm q}}} \lambda_i \ket{\psi_i}\bra{\psi_i},
\end{equation}
where $\ket{\psi_i}$ denotes the eigenvector associated with the $i$-th largest eigenvalue of $\rho$, and $\ket{\psi}=\ket{\psi_1}$ is the dominant eigenvector, which approximates the noise-free output state \cite{koczor,koczor_dominant_eigenvector}. Inspired by these observations, Koczor \cite{koczor} proposed the permutation-based quantum error mitigation concept (which has later been generalized to the concept of \ac{vd} \cite{permutation2}), as portrayed in Fig.~\ref{fig:koczor}. Compared to the permutation test shown in Fig.~\ref{fig:perm_test}, it may be observed that the output of the \ac{vd} circuit for a given unitary observable $\Sop{U}$ is given by
\begin{equation}\label{numerator_vd}
\tilde{y}_{\rm VD}^{(n)}=\tr{\rho^n\Sop{U}},
\end{equation}
where $n$ is the order of the circuit $\Sop{P}_n$, and we will also refer to it as the order of \ac{vd}. Another implementation yielding the same result as in \eqref{numerator_vd} is proposed in \cite{permutation2}, as shown in Fig.~\ref{fig:vd_another}. This implementation facilitates simultaneous measurement of multiple compatible observables, and thus reduces the total number of circuit repetitions. Note that all Pauli strings are unitary observables, hence they can be nicely fit into this framework. Next, upon replacing the observable $\Sop{U}$ by the identity operator $\Sop{I}$ (i.e., the original $n$-th order permutation test), one may also compute $\tr{\rho^n}$, and obtain the final result\footnote{The accuracy of this normalization procedure may be further improved by replacing $\tr{\rho^n}$ with $\lambda_1$. However, $\lambda_1$ is typically not known prior to the computation, and is also difficult to compute exactly from the observations. By contrast, $\tr{\rho^n}$ is readily obtainable by observing the identity operator.}
\begin{equation}\label{koczor_result}
y_{\rm VD}^{(n)} = \frac{\tilde{y}_{\rm VD}^{(n)}}{\tr{\rho^n}} =\frac{\tr{\rho^n\Sop{U}}}{\tr{\rho^n}}.
\end{equation}
Note that
\begin{equation}\label{pre_output_koczor}
\tilde{y}_{\rm VD}^{(n)} = \lambda_1^n\bra{\psi}\Sop{U}\ket{\psi}+ (1-\lambda_1)^n\sum_{i=2}^{2^{N_{\rm q}}} p_i^n\bra{\psi_i}\Sop{U}\ket{\psi_i},
\end{equation}
where $p_i=\lambda_i(1-\lambda_1)^{-1}$ satisfies $\sum_{i=2}^{N_{\rm q}}p_i=1$. When $\lambda_1$ is far larger than the other eigenvalues, it becomes clear from \eqref{pre_output_koczor} that the term $(1-\lambda_1)^n$ decreases much more rapidly with $n$ than $\lambda_1^n$. Hence the contribution of the undesired components $\ket{\psi_i},~i>1$ to the final computation result is substantially reduced by \ac{vd}.

\section{Permutation Filters}\label{sec:filter}
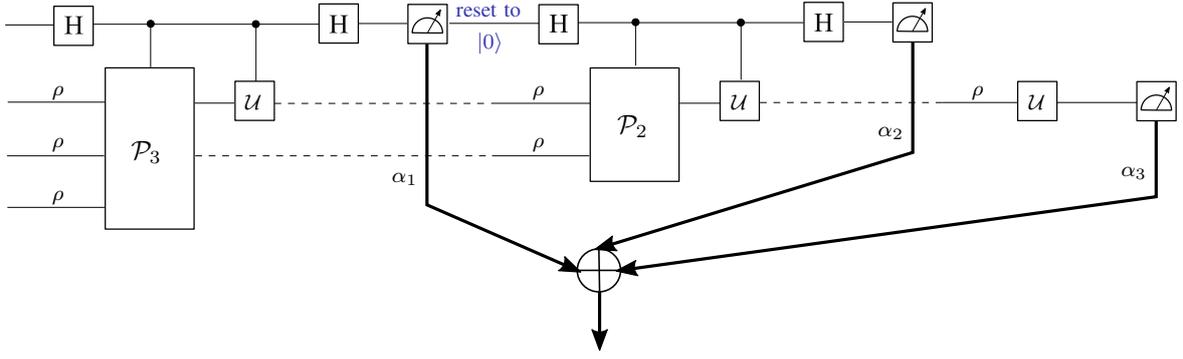
\begin{figure*}[t]
    \centering
   \begin{overpic}[width=.85\textwidth]{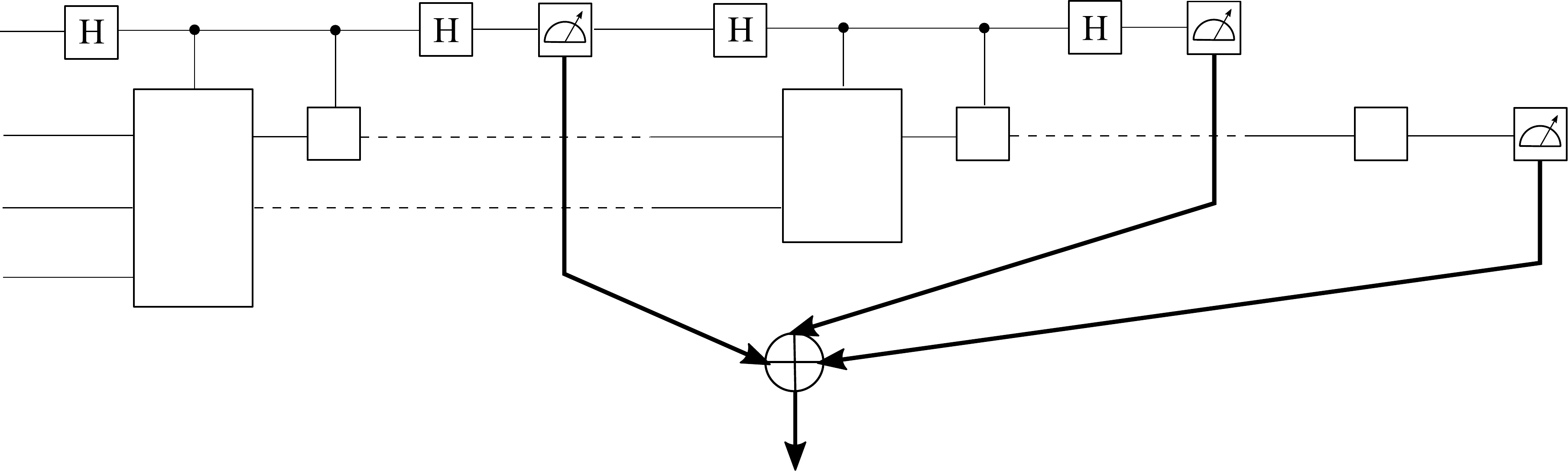}
   \put(4,22){\footnotesize $\rho$}
   \put(4,17.5){\footnotesize $\rho$}
   \put(4,13){\footnotesize $\rho$}
   \put(45,22){\footnotesize $\rho$}
   \put(45,17.5){\footnotesize $\rho$}
   \put(82.5,22){\footnotesize $\rho$}
   \put(10.75,16.5){$\mathcal{P}_3$}
   \put(52.25,18.75){$\mathcal{P}_2$}
   \put(95.25,15){\footnotesize $\alpha_3$}
   \put(74.5,18.5){\footnotesize $\alpha_2$}
   \put(33,14.5){\footnotesize $\alpha_1$}
   \put(20.5,20.8){\footnotesize $\Sop{U}$}
   \put(62,20.8){\footnotesize $\Sop{U}$}
   \put(87.4,20.8){\footnotesize $\Sop{U}$}
   \put(38.5,28.75){\footnotesize \color{myblue} reset to}
   \put(40.25,26){\footnotesize \color{myblue} $\ket{0}$}
   \end{overpic}
    \caption{Schematic of a third-order permutation filter ${\rm Tr}\{(\alpha_1\rho^3+\alpha_2\rho^2+\alpha_3\rho)\mathcal{U}\}$.}
    \label{fig:d_filter}
\end{figure*}

In this section, we propose a generalized version of virtual distillation, which will be referred to as ``permutation filters''. A third-order permutation filter is portrayed in Fig.~\ref{fig:d_filter}. As it may be observed from the figure, the third-order filter consists of the third-order and the second-order \ac{vd} circuits. In general, an $n$-th order permutation filter would contain all the $m$-th order \ac{vd} circuits, where $m=2,3,\dotsc,n$. Note that these circuits can be activated one after the other by reusing the same qubit resources, since the post-processing stage only involves a weighted averaging of the measured outcomes, which are classical quantities.

Formally, an $N$-th order permutation filter may be expressed as an $N$-th order polynomial of the input state $\rho$ formulated as
\begin{equation}\label{poly_filter_op}
\mathcal{F}_{\V{\alpha}}(\rho) = \sum_{n=1}^N \alpha_{N-n+1} \rho^n,
\end{equation}
where $\V{\alpha}=[\alpha_1~\alpha_2~\dotsc~\alpha_N]^{\rm T}\in\mathbb{R}^N$. Correspondingly, the eigenvalues of the output state are thus given by
\begin{equation}
h_{\V{\alpha}}(\lambda) = \sum_{n=1}^N \alpha_{N-n+1} \lambda^n.
\end{equation}
Observe that the function $h_{\V{\alpha}}(\lambda)$ may be viewed as the ``spectral response'' of the filter, resembling the frequency response of conventional filters used in classical signal processing tasks. The final computational result with respect to an observable $\Sop{U}$ is given by
\begin{equation}\label{final_computation}
y_{\rm filter}^{(N)}(\Sop{U})=\frac{\tr{\mathcal{F}_{\V{\alpha}}(\rho)\Sop{U}}}{\tr{\mathcal{F}_{\V{\alpha}}(\rho)}}.
\end{equation}

The reason that we do not include the constant term $\alpha_{N+1}$ in \eqref{poly_filter_op} is that it does not contribute to the final computational results in \eqref{final_computation} for most practical applications. To elaborate, consider the Pauli string decomposition \eqref{sum_string} of observables used in variational quantum algorithms. Since the single-qubit Pauli operators except for the identity have a trace of zero, we have $\tr{\Sop{U}}=0$ for every Pauli string $\Sop{U}$. Therefore, even if we include the constant coefficient $\alpha_{N+1}$ in our filter, it will not contribute to the final result, since we have:
\begin{equation}
\alpha_{N+1}\tr{\rho^0\Sop{U}}=0.
\end{equation}
As for the term involving the identity operator, we could simply account for it by adding a constant to the final computational result, since $\tr{\rho}=1$ always holds.

It is often convenient to design filters under an alternative parametrization, namely the pole-zero representation widely used in classical signal processing theory\footnote{In classical signal processing theory, filters are represented by a ratio between two polynomials in the complex frequency domain. The term ``poles'' refers to the roots of the denominator polynomial, while ``zeros'' refer to the roots of the numerator polynomial.}. When considering ``\ac{fir}-like'' filters taking the form \eqref{poly_filter_op} (since there is no denominator in this formula), there are only zeros but no poles. Observe from \eqref{poly_filter_op} that the first zero is at $\beta=0$ due to the lack of the constant term. Upon denoting the remaining zeros by $\V{\beta}=[\beta_1~\dotsc~\beta_{N-1}]^{\rm T}$, we have
\begin{equation}
\mathcal{F}_{\V{\beta}}(\rho)=\rho\prod_{n=1}^{N-1}(\rho-\beta_n\M{I}),
\end{equation}
and
\begin{equation}
h_{\V{\beta}}(\lambda)=\lambda\prod_{n=1}^{N-1} (\lambda-\beta_n).
\end{equation}
The relationship between $\V{\alpha}$ and $\V{\beta}$ is
\begin{equation}\label{def_alpha}
\V{\alpha} =\mathop{\bigstar}_{n=1}^{N-1} ~[1,~-\beta_n]^{\rm T},
\end{equation}
where we define $\mathop{\bigstar}_{n=1}^{K}\V{v}_n:=\V{v}_1\star \V{v}_2 \star \dotsc \star\V{v}_K$, and $\star$ denotes the discrete convolution given by
$$
[\V{x}\star\V{y}]_n=\sum_{i=\max\{1,k+1-n\}}^{\min\{k,m\}} x_i y_{k-i+1},
$$
where $\V{x}\in\mathbb{R}^m$, $\V{y}\in\mathbb{R}^n$, and $\V{x}\star\V{y} \in\mathbb{R}^{m+n-1}$. Without loss of generality, we assume that
\begin{equation}\label{order_zeros}
\beta_1\le\beta_2\le \dotsc \le \beta_{N-1}.
\end{equation}

\subsection{The Performance Metric of Permutation Filter Design}
For a given observable $\Sop{U}$, we would hope to minimize the estimation error
\begin{equation}
\begin{aligned}
\epsilon_{\Sop{U}}(\V{\beta})&=\left|y_{\rm filter}^{(N)}(\Sop{U})-\bra{\psi}\Sop{U}\ket{\psi}\right| \\
&=\left|\frac{\sum_{i=1}^{2^{N_{\rm q}}}h_{\V{\beta}}(\lambda_i)\bra{\psi_i}\Sop{U}\ket{\psi_i}}{\sum_{i=1}^{2^{N_{\rm q}}}h_{\V{\beta}}(\lambda_i)}-\bra{\psi}\Sop{U}\ket{\psi}\right|\\
&=\left|\frac{\frac{1}{h_{\V{\beta}}(\lambda_1)}\sum_{i=2}^{2^{N_{\rm q}}}h_{\V{\beta}}(\lambda_i)\left(\bra{\psi_i}\Sop{U}\ket{\psi_i}-\bra{\psi}\Sop{U}\ket{\psi}\right)}{1+[h_{\V{\beta}}(\lambda_1)]^{-1}\sum_{i=2}^{2^{N_{\rm q}}}h_{\V{\beta}}(\lambda_i)}\right|.
\end{aligned}
\end{equation}
However, in a typical variational quantum algorithm, a large number of unitary observables $\Sop{U}_1,\dotsc,\Sop{U}_{N_{\rm ob}}$ would have to be evaluated. In light of this, we consider the minimization of the following upper bound
\begin{equation}
\begin{aligned}
\epsilon_{\Sop{U}}(\V{\beta})&\le \epsilon(\V{\beta})\\
&=\frac{2}{h_{\V{\beta}}(\lambda_1)}\left\|\V{h}_{\V{\beta}}(\tilde{\V{\lambda}})\right\|_1,
\end{aligned}
\end{equation}
where $\tilde{\V{\lambda}}=[\V{\lambda}]_{2:2^{N_{\rm q}}}$, and $\V{\lambda}=[\lambda_1~\dotsc~\lambda_{2^{N_{\rm q}}}]^{\rm T}$.

If we know \textit{a priori} the distribution of $\tilde{\V{\lambda}}$, or in other words, the spectral density of $\rho$ (excluding the dominant eigenvalue), we may directly minimize the cost function $\epsilon(\V{\beta})$ as follows:
\begin{equation}\label{optimization_problem}
\begin{aligned}
\min_{\V{\beta}} &~~ \epsilon(\V{\beta}), ~~\mathrm{s.t.}~~\V{\beta}\in\Set{B},~\eqref{order_zeros},
\end{aligned}
\end{equation}
where $\epsilon(\V{\beta})$ can be rewritten as
$$
\epsilon(\V{\beta})=\frac{1}{\lambda_1\prod_{n=1}^{N-1}(\lambda_1-\beta_n)} \int_{\lambda_{\rm m}}^1 \left|\lambda\prod_{n=1}^{N-1}(\lambda-\beta_n)\right|   f(\lambda) {\rm d}\lambda,
$$
$\lambda_{\rm m}>0$ denotes the minimum value of $\lambda$, and $f(\lambda)$ denotes the spectral density. The feasible region $\Set{B}$ is given by
$$
\Set{B}=\{\V{\beta}|\V{\beta}\succeq \V{0},\beta_1\le\beta_2\le\dotsc \le\beta_{N-1}\}.
$$
For most practical scenarios, we have $\beta_i\ll \lambda_1$, hence $\epsilon(\V{\beta})$ may be approximated as
\begin{equation}\label{approx_beta}
\epsilon(\V{\beta})\approx\tilde{\epsilon}(\V{\beta})=\int_{\lambda_{\rm m}}^1 \left|\lambda\prod_{n=1}^{N-1}(\lambda-\beta_n)\right|   f(\lambda) {\rm d}\lambda,
\end{equation}
since the optimal solution is hardly affected by the denominator. Note that $h_{\V{\beta}}(1)$ is always positive, hence we may further simplify the approximated objective function as follows:
\begin{equation}\label{obj_pareto}
\begin{aligned}
\tilde{\epsilon}(\V{\beta})&=\int_{\lambda_{\rm m}}^1 \left|G_{\V{\beta}}(\lambda)\right| {\rm d}\lambda\\
&=\sum_{i=0}^{N-1}(-1)^i\int_{\beta_{N-i-1}}^{\beta_{N-i}}G_{\V{\beta}}(\lambda) {\rm d}\lambda,
\end{aligned}
\end{equation}
where $G_{\V{\beta}}(\lambda)=f(\lambda)\lambda\prod_{n=1}^{N-1}(\lambda-\beta_n)$, and additionally we define $\beta_N=1$ and $\beta_0=\lambda_{\rm m}$.

\subsection{Practical Permutation Filter Design Algorithms}
When $f(\lambda)$ is known exactly, we may directly solve the optimization problem discussed in the previous subsection. However, for practical applications, $f(\lambda)$ is never known precisely; it has to be estimated from observations. In this treatise, we fit Pareto distribution \cite{pareto1,pareto2} to $f(\lambda)$ which is formulated as:
\begin{equation}
f(\lambda)=k\lambda_{\rm m}^k\lambda^{-(k+1)},
\end{equation}
where $k>2$ is a shape parameter.

The reason for using the Pareto distribution is two-fold. First of all, it approximates our empirical observations concerning the output spectra of noisy quantum circuits quite closely. Secondly, it fits nicely with the polynomial form of the permutation filter, making the design problem more tractable. Specifically, under the parametrization of the Pareto distribution, the indefinite integral of $G_{\V{\beta}}(\lambda)$ can be explicitly calculated as follows:
\begin{equation}\label{alpha_coefficients}
\begin{aligned}
\tilde{G}_{\V{\alpha}}(\lambda)&=\frac{1}{k\lambda_{\rm m}^k}\int G_{\V{\beta}}(\lambda) {\rm d}\lambda \\
&=\int \lambda^{-k}\prod_{n=1}^{N-1}(\lambda-\beta_n) {\rm d}\lambda =\sum_{n=1}^N \frac{\alpha_{N-n+1}}{n-k}\cdot \lambda^{n-k}.
\end{aligned}
\end{equation}
The definite integrals in \eqref{obj_pareto} can then be obtained as
\begin{equation}\label{definite_integral}
\int_{\beta_i}^{\beta_{i+1}} G_{\V{\beta}}(\lambda){\rm d}\lambda = k\lambda_{\rm m}^k\left(\tilde{G}_{\V{\alpha}}(\beta_{i+1})-\tilde{G}_{\V{\alpha}}(\beta_i)\right).
\end{equation}

Note that for an $N$-th order permutation filter, we may obtain $N-1$ observations $\V{m}=[m_1~\dotsc~m_{N-1}]^{\rm T}$ where $m_i=\tr{\rho^{i+1}}$. These observations can be used to fit the Pareto distribution to $f(\lambda)$ using the method of moments \cite{kay}. For example, when $N=3$, the equations of moments are given by
\begin{equation}\label{eq_moments}
\begin{aligned}
\frac{1-\hat{\lambda}_1(\V{m})}{2^{N_{\rm q}}-1}&=\frac{k\lambda_{\rm m}}{k-1}, \\
\frac{m_1-\hat{\lambda}_1(\V{m})^2}{2^{N_{\rm q}}-1}&=\frac{k\lambda_{\rm m}^2}{k-2},
\end{aligned}
\end{equation}
where $\hat{\lambda}_1(\V{m})$ is an estimate of $\lambda_1$. Here, the quantities $\frac{1-\hat{\lambda}_1(\V{m})}{2^{N_{\rm q}}-1}$ and $\frac{m_1-\hat{\lambda}_1(\V{m})^2}{2^{N_{\rm q}}-1}$ are estimates of the mean value and the variance of the spectrum, respectively. We do not use the conventional sample mean and variance, because the eigenvalues cannot be sampled directly.  A natural choice of $\hat{\lambda}_1(\V{m})$ for an $N$-th order filter is
\begin{equation}
\hat{\lambda}_1(\V{m}) = \|\V{\lambda}\|_N=(m_{N-1})^{\frac{1}{N}},
\end{equation}
which is asymptotically exact as $N\rightarrow \infty$, since $\lambda_1 = \|\V{\lambda}\|_{\infty}$.

Using the equations of moments in \eqref{eq_moments}, we may then estimate the unknown parameters $k$ and $\lambda_{\rm m}$. However, for the $N=2$ case, the method of moments would encounter an identifiability problem, since the number of observations (one) is less than the number of parameters (two). Fortunately, we may obtain the closed-form solution of $\beta_1$ as follows:
\begin{equation}\label{closed_form1}
\beta_1=\lambda_{\rm m}\left(\frac{2}{1+\lambda_{\rm m}^{k-1}}\right)^{\frac{1}{k-1}},
\end{equation}
which is obtained by taking the derivative of $\tilde{\epsilon}(\V{\beta})$ with respect to $\beta_1$ and setting it to zero. For $k\ge 2$, $\beta_1$ can be closely approximated by
\begin{equation}\label{closed_form2}
\beta_1\approx \mu=k\lambda_{\rm m}(k-1)^{-1},
\end{equation}
where $\mu$ is the mean value of the Pareto distribution. This may be seen by neglecting the term $\lambda_{\rm m}^{k-1}$ (since typically $\lambda_{\rm m}^{k-1}\ll 1$ when $k\ge 2$), and noticing that the ratio $\mu/\beta_1$ is then approximately (approximately because of neglecting $\lambda_{\rm m}^{k-1}$) bounded by

$$
1\lessapprox \mu/\beta_1 \lessapprox \frac{2^{-\frac{1-\ln 2}{\ln 2}}}{\ln 2} \approx 1.062,
$$
where the lower bound is attained at $k=2$ and the upper bound is attained at $k=(1-\ln 2)^{-1}$. The mean value $\mu$ may then be estimated by
\begin{equation}\label{closed_form3}
\hat{\mu} = \frac{1-\hat{\lambda}_1(\V{m})}{2^{N_{\rm q}}-1}.
\end{equation}

For the $N>2$ case, it is difficult to obtain closed-form solutions of $\V{\beta}$. Furthermore, in general, the optimization problem with respect to $\V{\beta}$ may no longer be convex. Fortunately, in the following proposition we show that $\tilde{\epsilon}(\V{\beta})$ satisfies a generalized convexity property, which guarantees that the global optimum is always attainable.
\begin{proposition}[Invexity of the Permutation Filter Design Problem]\label{prop:global_minimum}
The cost function $\tilde{\epsilon}(\V{\beta})$ in \eqref{optimization_problem} is an invex\footnote{Invexity is a generalization of convexity, ensuring that the global optimal solutions can be found by using the Karush-Kuhn-Tucker conditions \cite{invex2}.} function of $\V{\beta}$ in the convex feasible region $\Set{B}$. In other words, every stationary point of $\tilde{\epsilon}(\V{\beta})$ in $\Set{B}$ is a global minimum.
\begin{IEEEproof}
Please refer to Appendix \ref{sec:proof_global_minimum}.
\end{IEEEproof}
\end{proposition}

Proposition \ref{prop:global_minimum} implies that the following simple projected gradient descent iteration rule
\begin{equation}
\begin{aligned}
\tilde{\V{\beta}}^{(\ell+1)}&=\V{\beta}^{(\ell)}-\delta^{(\ell)} \cdot \left.\frac{\partial \tilde{\epsilon}(\V{\beta})}{\partial \V{\beta}}\right|_{\V{\beta}^{(\ell)}}, \\
\V{\beta}^{(\ell+1)}&=\Sop{T}_{\Set{B}}\left[\tilde{\V{\beta}}^{(\ell+1)}\right],
\end{aligned}
\end{equation}
may be used to solve the problem in \eqref{optimization_problem}, despite that $\tilde{\epsilon}(\V{\beta})$ may not be convex with respect to $\V{\beta}$. The operator $\Sop{T}_{\Set{B}}(\cdot)$ projects its argument onto the convex feasible region $\Set{B}$, which can be implemented by simply sorting the entries of $\V{\beta}$ after each iteration. The step size parameter $\delta^{(l)}$ can be determined using classic line search methods \cite{line_search}. More sophisticated methods, such as modified Newton's method specifically tailored for invex optimization \cite{newtonmr}, may also be applied to accelerate the convergence.

According to our discussion in Appendix \ref{sec:proof_global_minimum}, the cost function $\xi(\V{\alpha})$ is a convex function of $\V{\alpha}$. The reason that we do not solve directly this convex problem is that it is a challenge to differentiate the cost function $\xi(\V{\alpha})$. By contrast, it is relatively simple to compute the gradient $\frac{\partial}{\partial \V{\beta}}\tilde{\epsilon}(\V{\beta})$, as follows:
\begin{equation}\label{gradient_main}
\begin{aligned}
\frac{\partial}{\partial \V{\beta}}\tilde{\epsilon}(\V{\beta}) &= \sum_{i=0}^{N-1}(-1)^i\frac{\partial}{\partial \V{\beta}}\int_{\beta_{N-i-1}}^{\beta_{N-i}}G_{\V{\beta}}(\lambda) {\rm d}\lambda \\
&= \sum_{i=0}^{N-1}(-1)^i\int_{\beta_{N-i-1}}^{\beta_{N-i}}\frac{\partial}{\partial \V{\beta}}G_{\V{\beta}}(\lambda) {\rm d}\lambda \\
&=\sum_{i=0}^{N-1}(-1)^{i+1}\int_{\beta_{N-i-1}}^{\beta_{N-i}}\V{g}_{\V{\beta}}(\lambda) {\rm d}\lambda,
\end{aligned}
\end{equation}
where $[\V{g}_{\V{\beta}}(\lambda)]_i=\lambda^{-k}\prod_{\substack{n=1\\n\neq i}}^{N-1}(\lambda-\beta_n)$. The order between the integration and the differentiation is interchangeable, since $G_{\V{\beta}}(\lambda)=0$ for $\lambda = \beta_i,~\forall i=1,2,\dotsc,N-1$. The integrals can be computed using \eqref{alpha_coefficients} and \eqref{definite_integral}, but for $[\V{g}_{\V{\beta}}(\lambda)]_i$ the vector $\V{\alpha}$ should be replaced by
\begin{equation}\label{modified_convolution}
\tilde{\V{\alpha}}_i=\mathop{\bigstar}_{n=1,n\neq i}^{N-1}~ [1~-\beta_n]^{\rm T}.
\end{equation}

{
\begin{algorithm}[t]
 \caption{Type-2 permutation filter design}
 \label{alg:type_2_design}
 \begin{algorithmic}[1]
 \renewcommand{\algorithmicrequire}{\textbf{Input:}}
 \renewcommand{\algorithmicensure}{\textbf{Output:}}
 \REQUIRE Spectral density parameters $k$ and $\lambda_{\rm m}$
 \ENSURE The filter weight vector $\V{\alpha}$
  \STATE $\ell=0$; Initialize $\V{\beta}^{(0)}$;
  \REPEAT
  \STATE Compute $\left.\frac{\partial \tilde{\epsilon}(\V{\beta})}{\partial \V{\beta}}\right|_{\V{\beta}^{(\ell)}}$ using \eqref{alpha_coefficients}, \eqref{definite_integral}, \eqref{gradient_main} and \eqref{modified_convolution};
  \STATE Determine $\delta^{(\ell)}$ using line search methods;
  \STATE Update $\V{\beta}^{(\ell+1)}=\V{\beta}^{(\ell)}-\delta^{(\ell)} \cdot \left.\frac{\partial \tilde{\epsilon}(\V{\beta})}{\partial \V{\beta}}\right|_{\V{\beta}^{(\ell)}}$;
  \STATE Sort the entries in $\V{\beta}^{(\ell+1)}$ in the ascending order;
  \STATE $\ell=\ell+1$;
  \UNTIL{convergence conditions are met}
  \STATE Compute $\V{\alpha}=\V{\varphi}(\V{\beta}^{(\ell)})$ using \eqref{def_alpha};
 \RETURN $\V{\alpha}$
 \end{algorithmic}
 \end{algorithm}

  }

When low-complexity methods are preferred, a simple heuristic alternative, which will be referred to as the ``Type-1 permutation filter'', is to set
\begin{equation}
\beta_1=\beta_2=\dotsc=\beta_{N-1}=\mu.
\end{equation}
Correspondingly, we refer to the aforementioned optimization-based method, summarized in Algorithm \ref{alg:type_2_design}, as the ``Type-2 permutation filter''. In Section~\ref{sec:performance} we will show that, even though the Type-1 filters rely on a heuristic method, they are capable of outperforming \ac{vd}.

To conclude, the complete workflow of an $N$-th order permutation filter for a given observable $\Sop{U}$ consists of the following steps:
\begin{enumerate}
\item Execute the original circuit and obtain the estimate of $\tr{\rho \Sop{U}}$;
\item Execute all $n$-th order virtual distillation circuits ($2\le n \le N$), and obtain the estimates of $\tr{\rho^n\Sop{U}}$ as well as additional observations $m_{n-1}=\tr{\rho^n}$;
\item Fit the spectral density model using the observations $\V{m}=[m_1~\dotsc~m_{N-1}]^{\rm T}$, and determine the filter parameters $\V{\alpha}$;
\item Obtain the final filtered result by classical post-processing.
\end{enumerate}
\subsection{The Computational Overhead of Permutation Filters}
In terms of the number of gates, the computational overhead of permutation filters is the same as that of virtual distillation. The number of gates required for implementing the permutation operation $\Sop{P}_n$ (which is the additional gate cost of the protocol compared to the unprotected circuit) has been discussed in \cite{koczor}. Specifically, if the original unprotected circuit acts on $N_{\rm q}$ qubits, implementing $\Sop{P}_n$ would require $N_{\rm q}(n-1)$ controlled-SWAP gates (i.e. the Fredkin gate), which is on the order of $O(N_{\rm q})$. Hence we may conclude that the method would be beneficial, when the unprotected circuit has an increasing depth with respect to $N_{\rm q}$.

As for the sampling overhead, permutation filters are slightly different from virtual distillation due to the weighted averaging process. For virtual distillation, an approximate expression for the variance of a given observable $\Sop{U}$ has been presented in \cite{permutation2}. Using similar arguments, we may also obtain an expression for permutation filters formulated as
\newcommand{\mB}{\alpha_N+\sum_{n=2}^N\alpha_{N-n+1}\tr{\rho^n}}
\newcommand{\vB}{1\!-\!\sum_{n=2}^N \alpha_{N-n+1}^2\tr{\rho^n}^2}
\newcommand{\mA}{\sum_{n=1}^N\alpha_{N-n+1}\tr{\rho^n\Sop{U}}}
\newcommand{\vA}{1\!-\!\sum_{n=1}^N \alpha_{N-n+1}^2 \tr{\rho^n\Sop{U}}^2}
\newcommand{\covAB}{\sum_{n=2}^N\alpha_{N-n+1}^2\left(\tr{\rho\Sop{U}}-\tr{\rho^n\Sop{U}}\tr{\rho^n}\right)}
\begin{equation}
\begin{aligned}
&\hspace{3mm}{\rm Var}\{y_{\rm filter}^{(N)}(\Sop{U})\} \\
&\approx \frac{\vA}{(\mB)^2} \\
&-\frac{2(\mA)}{(\mB)^3} \\
&\times  \covAB \\
&+\!\frac{(\mA)^2(\vB)}{(\mB)^4}.
\end{aligned}
\end{equation}
The variance of the entire Hamiltonian $\Sop{H}$ may then be calculated by a weighted summation over the Pauli observables. In light of this, the sampling overhead factor of permutation filters may be defined as the ratio between the variance of the Hamiltonian estimator based on the permutation filter and that based on the unprotected circuit. We will evaluate the sampling overhead of permutation filters applied to practical variational quantum algorithms using this metric in Section \ref{ssec:qaoamud}.

\section{The Error Reduction Performance of Permutation Filters}\label{sec:performance}
In this section, we quantify the error reduction of permutation filters compared to \ac{vd} of the same order using the following performance metric.
\begin{definition}[Error Ratio]
We define the error ratio between an $N$-th order permutation filter $\mathcal{F}_{\V{\beta}}(\cdot)$ and its corresponding $N$-th order counterpart based on \ac{vd} as follows:
\begin{equation}\label{def_ratio}
R(\V{\beta}):=\frac{\tilde{\epsilon}(\V{\beta})}{\tilde{\epsilon}(\V{0})}.
\end{equation}
Note that \ac{vd} is equivalent to a permutation filter that satisfies $\V{\beta}=\V{0}$.
\end{definition}

\begin{figure}
\centering
\includegraphics[width=.5\textwidth]{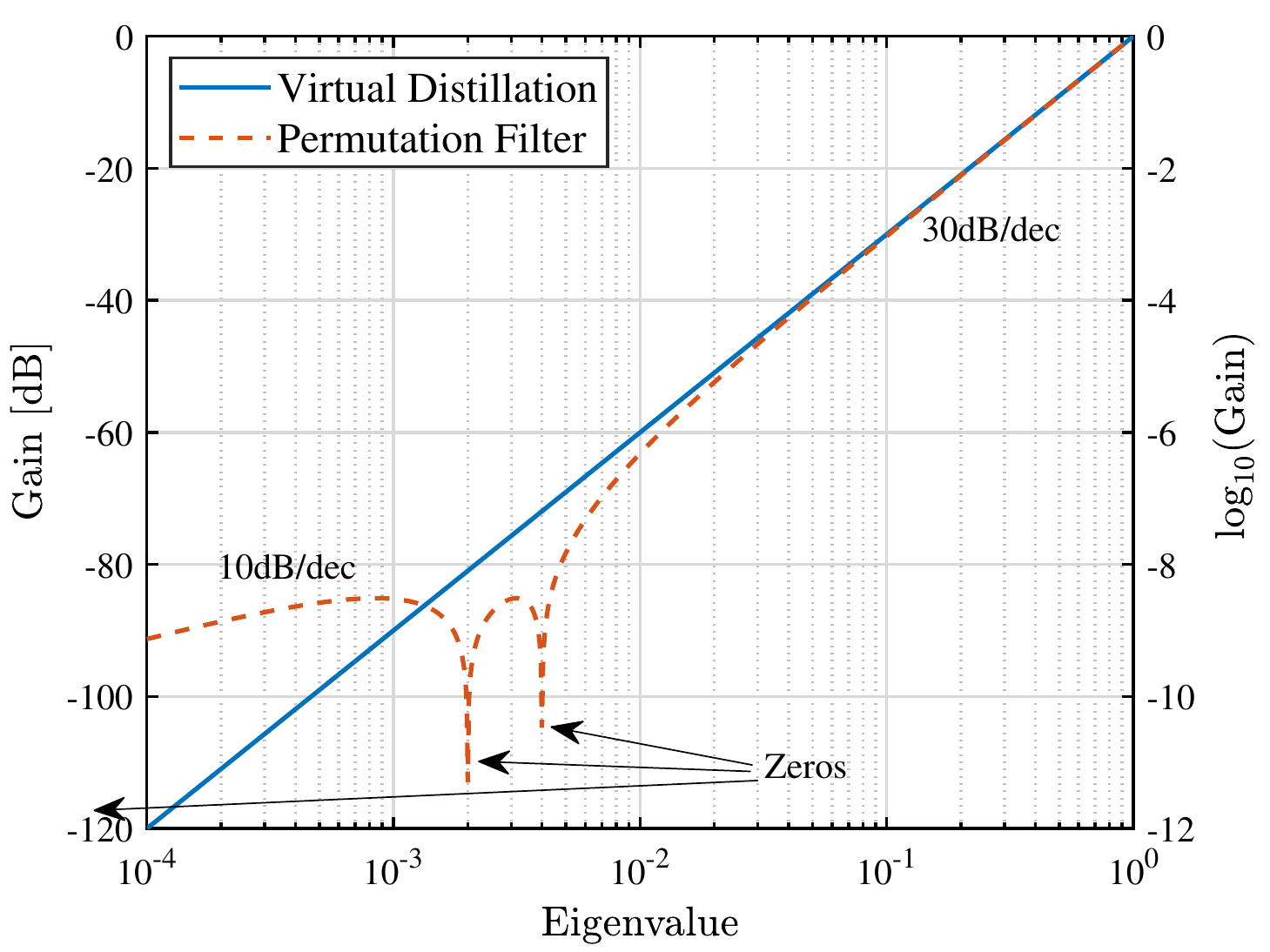}

\caption{The spectral response of a third-order permutation filter, compared to that of the third-order \ac{vd}.}
\label{fig:illustration_filter}

\end{figure}

Intuitively, the permutation filters are narrowband notch filters, hence they should perform better when the ``bandwidth'' of the undesired spectral components is lower. To see this more clearly, we consider the spectral response of a third-order permutation filter, as portrayed in Fig.~\ref{fig:illustration_filter}. Observe that every zero contributes $10$ dB per decade to the slope of the filter gain.\footnote{For readers do not familiar with classical signal processing theory, please refer to Appendix \ref{sec:notes_spectral_response} for further explanation.} For both third-order permutation filters and for \ac{vd}, the slope will be $30$ dB per decade beyond the largest zero. In light of this, the only region where permutation filters have smaller gain is the narrowband range around the two largest zeros. Therefore, permutation filters perform the best when the noise components are concentrated in this region.

To make our aforementioned intuitions more rigorous, we define the following quantities to characterize the bandwidth.
\begin{definition}[Noise Bandwidth]
We define the bandwidth of the noise (i.e., the undesired spectral components $\tilde{\V{\lambda}}$ in a mixed state $\rho$) as follows:
\begin{equation}
B(\tilde{\V{\lambda}}):=\sqrt{\mathbb{E}\{|\lambda-\mu|^2\}},
\end{equation}
where
\begin{equation}
\mathbb{E}\{g(\lambda)\}:=\int_{\lambda_{\rm m}}^1 g(\lambda)f(\lambda){\rm d}\lambda,
\end{equation}
denotes the expectation operation, and $\mu=\mathbb{E}\{\lambda\}$ denotes the mean value of noise components. We also define the \textit{relative noise bandwidth} as
\begin{equation}
b(\tilde{\V{\lambda}}):=\mu^{-1} B(\tilde{\V{\lambda}}).
\end{equation}
\end{definition}

Given the previous definitions, we are now prepared to state the following result concerning the error ratio of Type-1 permutation filters.
\begin{proposition}[Generic Error Ratio Scaling Behaviour of Type-1 Permutation Filters]\label{prop:main}
The error ratio $R(\V{\beta})$ of an $N$-th order Type-1 permutation filter, as a function of the relative noise bandwidth $b(\tilde{\V{\lambda}})$, can be bounded by
\begin{equation}\label{scaling_desired}
R(\V{\beta})\le \frac{1}{\mu}\left[b(\tilde{\V{\lambda}})\sqrt{2^{N_{\rm q}}-1}\right]^{N-1},
\end{equation}
as $b(\tilde{\V{\lambda}})\rightarrow 0$.
\begin{IEEEproof}
The term $\tilde{\epsilon}(\V{0})$ can thus be written explicitly as
\begin{equation}
\tilde{\epsilon}(\V{0}) = \mathbb{E}\{\lambda^N\}.
\end{equation}
Using Jensen's inequality \cite{convex_opt}, we have
\begin{equation}
\tilde{\epsilon}(\V{0})\ge [\mathbb{E}\{\lambda\}]^N = \mu^N.
\end{equation}
Therefore, from \eqref{def_ratio} we obtain
\begin{equation}
\begin{aligned}
R(\V{\beta})&\le \tilde{\epsilon}(\V{\beta})\mu^{-N}\\
&=\mathbb{E}\{|\lambda(\lambda-\mu)^{N-1}|\}\mu^{-N} \\
&\le \frac{1}{\mu}\cdot \mathbb{E}\left\{\left|(\lambda-\mu)\mu^{-1}\right|^{N-1}\right\},
\end{aligned}
\end{equation}
where the last line follows from the fact that $\lambda\le 1$ holds for all eigenvalues. Furthermore, assume that we have access to the actual values of $\tilde{\V{\lambda}}$ (which will only be used for calculating intermediate results), we have

\begin{equation}
\begin{aligned}
\mathbb{E}\left\{\left|\frac{\lambda-\mu}{\mu}\right|^{N-1}\right\}&=\left(\frac{\mu^{-1}\left\|\tilde{\V{\lambda}}-\mu\V{1}\right\|_{N-1}}{(2^{N_{\rm q}}-1)^{\frac{1}{N-1}}}\right)^{N-1}\\
&\le \left(\mu^{-1}\left\|\tilde{\V{\lambda}}-\mu\V{1}\right\|_{\infty}\right)^{N-1} \\
&\le \left(\mu^{-1}\left\|\tilde{\V{\lambda}}-\mu\V{1}\right\|_2\right)^{N-1} \\
&=\left(b(\tilde{\V{\lambda}})\sqrt{2^{N_{\rm q}}-1}\right)^{N-1}.
\end{aligned}
\end{equation}
Hence the proof is completed.
\end{IEEEproof}
\end{proposition}

Proposition \ref{prop:main} supports our intuition that the error ratio decreases, as the noise bandwidth becomes smaller. However, the constant $\sqrt{2^{N_{\rm q}}-1}$ in \eqref{scaling_desired} can be extremely large for large $N_{\rm q}$, when the bound becomes of limited practical significance. In the following result we show that for spectral densities satisfying Pareto distributions, the dependence of the bound on $N_{\rm q}$ can be eliminated.
\begin{proposition}[Type-1 Filters Applied to Pareto-Distributed States]\label{prop:pareto}
Assume that $f(\lambda)$ corresponds to a Pareto distribution, and that $b(\tilde{\V{\lambda}})<(N-1)^{-1}$. The error ratio of an $N$-th order Type-1 permutation filter can be bounded by
\begin{equation}\label{desired_pareto}
\begin{aligned}
R(\V{\beta})&\le \frac{(N-1)![1+b(\tilde{\V{\lambda}})]^N}{e\prod_{n=1}^{N-2}[1-nb(\tilde{\V{\lambda}})]} \cdot [b(\tilde{\V{\lambda}})]^{N-1}\\
&=O\left\{[b(\tilde{\V{\lambda}})]^{N-1}\right\}.
\end{aligned}
\end{equation}
\begin{IEEEproof}
Please refer to Appendix \ref{sec:proof_pareto}.
\end{IEEEproof}
\end{proposition}

Both Proposition \ref{prop:main} and \ref{prop:pareto} show that, the error ratio of Type-1 filters decreases exponentially with the filter order $N$. For Type-2 filters, this may be viewed as an upper bound of the error ratio, since their parameter vectors $\V{\beta}$ are obtained via optimization. By contrast, the parameter vectors of Type-1 filters are determined using only the mean value of noise components, hence are suboptimal.

A natural question that arises is: under what practical conditions do the undesired spectral components have small relative bandwidth? In the following proposition, we show that the relative noise bandwidth decreases with the depth of quantum circuits, as well as with the error rate of the gates in the circuits.
\begin{proposition}[Exponential Spectral Concentration of Deep Quantum Circuits]\label{prop:concentration}
Assume that each qubit is acted upon by at least $L$ gates, and that each of the gates is contaminated by quantum channels containing Pauli noise, which have matrix representations under the Pauli basis given in \eqref{pauli_channel_ptm}. We assume furthermore that the probability of each type of Pauli error (i.e., X error, Y error or Z error) on each qubit is lower bounded by $\epsilon_{\rm l}$. Under these assumptions, the relative noise bandwidth can be upper bounded by
\begin{equation}\label{exponential_concentration}
\begin{aligned}
b(\tilde{\V{\lambda}}) &\le \frac{1+\sqrt{2^{N_{\rm q}}-1}}{1-2^{-N_{\rm q}}-\exp(-4\epsilon_{\rm l}L)} \cdot \exp(-4\epsilon_{\rm l}L)\\
&=O\left\{\exp(-4\epsilon_{\rm l}L)\right\}.
\end{aligned}
\end{equation}
\begin{IEEEproof}
Please refer to Appendix \ref{sec:proof_concentration}.
\end{IEEEproof}
\end{proposition}

From Proposition \ref{prop:concentration} we observe that the relative noise bandwidth decreases exponentially with the product of $\epsilon_{\rm l}$ and $L$. This implies that the proposed permutation filters would provide more significant performance improvements when the circuits are relatively deep, or the gates therein are noisy.

\section{Numerical Results}\label{sec:numerical}
\begin{figure}
\centering
\includegraphics[width=.48\textwidth]{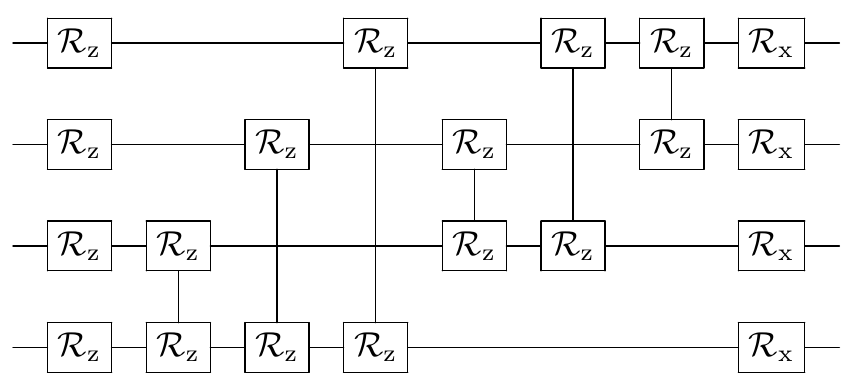}

\caption{Schematic of a stage in the parametric state preparation circuit used in the simulations. Here we set $N_{\rm q}=4$ only for illustration. }
\label{fig:stage}

\end{figure}

In this section, we further illustrate the results discussed in the previous sections using numerical simulations. In all simulations, we consider a class of parametric state preparation circuit consisting of different number of stages, for which a single stage is portrayed in Fig.~\ref{fig:stage}. For illustration we drawn a four-qubit circuit, but in the actual simulations we set $N_{\rm q}=10$. As observed from Fig.~\ref{fig:stage}, each stage of the circuit is constructed by two-qubit ZZ-rotation gates acting upon each pair of qubits, and single-qubit X- and Y-rotation gates acting upon each qubit. The rotation angle of each gate is a parameter to be determined. In the simulations, we choose the parameters by independent sampling from uniform distributions over $[-\pi,\pi]$, and the simulation results are averaged over $100$ random instances of the circuits. The gates are inflicted by depolarizing errors occurring at varying probabilities, but we always set the depolarizing probabilities of two-qubit gates $10$ times higher than that of single-qubit gates.

\subsection{Spectral Properties of the Output States}\label{ssec:spectral_property}
\begin{figure}[t]
\centering
\subfloat[][The spectral density]{
\centering
\includegraphics[width=.46\textwidth]{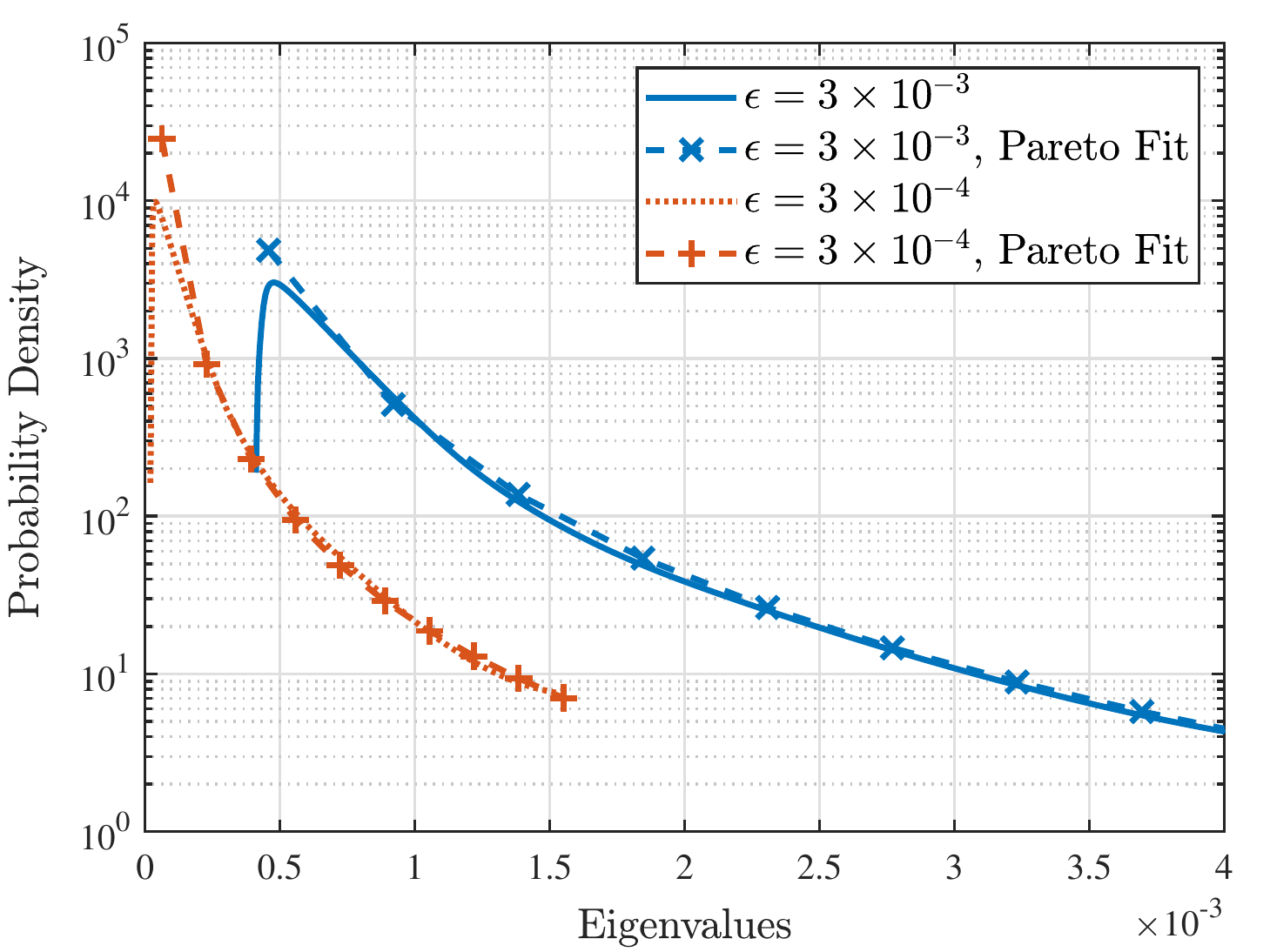}
\label{fig:spectra_pdf}
}
\\
\subfloat[][The cumulative distribution function]{
\centering
\includegraphics[width=.46\textwidth]{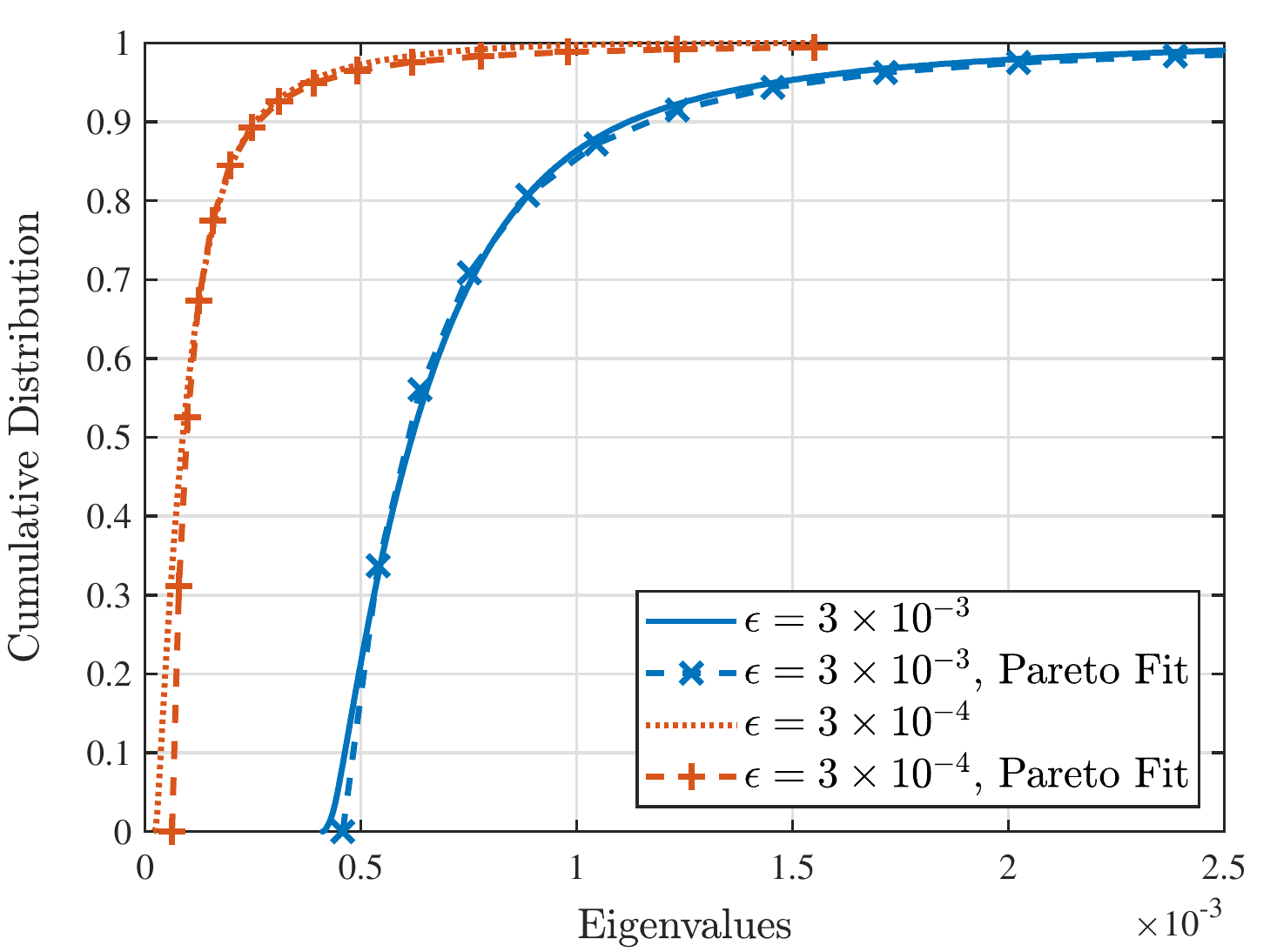}
\label{fig:spectra_cdf}
}
\caption{The spectra and the corresponding Pareto fits of the output states of parametric state preparation circuits having different depolarizing probability $\epsilon$.}
\label{fig:spectra_pareto}
\end{figure}

We first demonstrate the spectral densities of the output states. In particular, we consider parametric state preparation circuits having $10$ stages acting on $N_{\rm q}=10$ qubits. The spectral densities and the corresponding cumulative density functions for $\epsilon=3\times 10^{-4}$ and $\epsilon=3\times 10^{-3}$ are portrayed in Fig. \ref{fig:spectra_pareto}, where $\epsilon$ denotes the depolarizing probability of each two-qubit gate. The Pareto fit are also plotted for comparison. We see that the Pareto distributions provide good approximations to the eigenvalue spectra, except for very small eigenvalues. This also suggests that the Pareto fit may become less accurate when the noise bandwidth is very narrow, for which the approximation error becomes more significant.

\subsection{The Filter Design Metric $\tilde{\epsilon}(\V{\beta})$}\label{ssec:abs_error}
Next, we investigate the values of the cost function $\tilde{\epsilon}(\V{\beta})$ for filter design under different scenarios, which may be used for evaluating the performance of the filters irrespective of the specific choices of observables.

\begin{figure}
\centering
\subfloat[][Two-qubit depolarizing probability $1.25\times 10^{-3}$]{
\centering
\includegraphics[width=.46\textwidth]{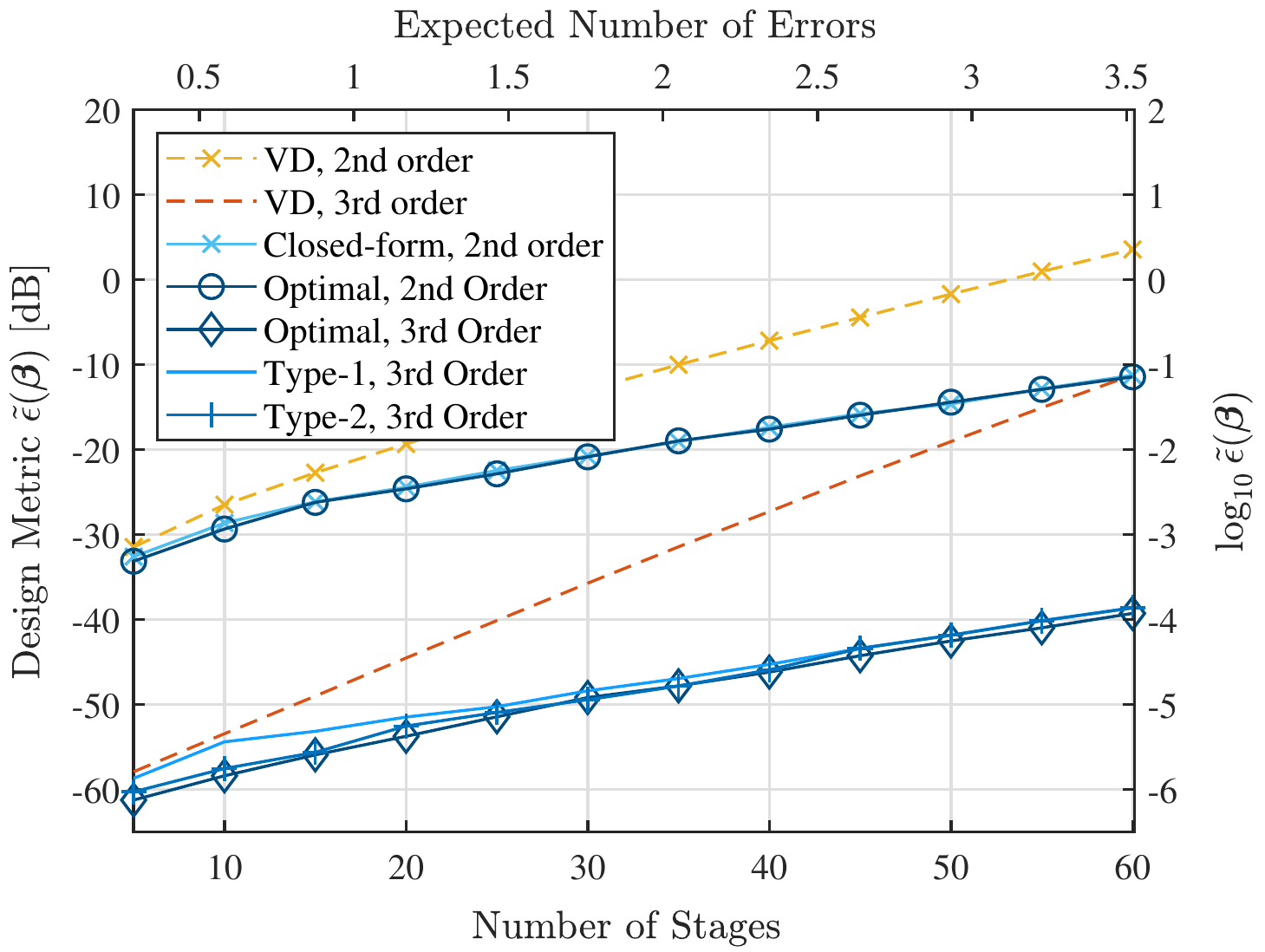}
\label{fig:error_ub_low}
}
\\
\subfloat[][Two-qubit depolarizing probability $5\times 10^{-3}$]{
\centering
\includegraphics[width=.46\textwidth]{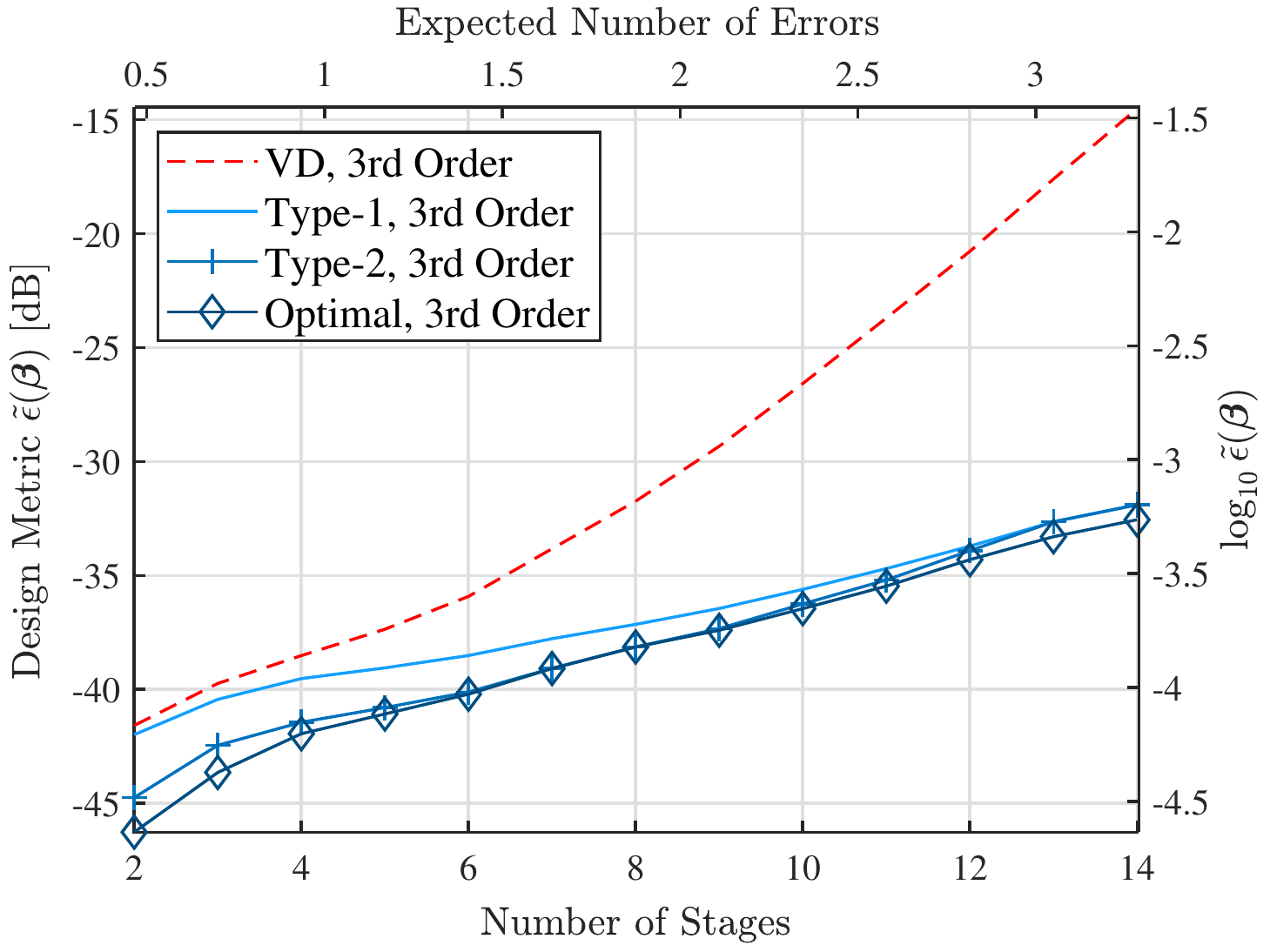}
\label{fig:error_ub_high}
}
\caption{The value of the design metric $\tilde{\epsilon}(\V{\beta})$ in \eqref{obj_pareto} for both \ac{vd} and for the proposed methods, as functions of the number of stages.}
\end{figure}

In Fig.~\ref{fig:error_ub_low}, we compare the values of $\tilde{\epsilon}(\V{\beta})$ obtained both by our permutation filters and by \ac{vd}, as functions of the number of stages in the state preparation circuits. The depolarizing probability of two-qubit gates is $1.25\times 10^{-3}$. In this figure, the curve ``Closed-form, 2nd order'' corresponds to the second-order permutation filter designed based on the closed-form solution in \eqref{closed_form1}--\eqref{closed_form3}. We observe from the figure that permutation filters significantly outperform \ac{vd}, when the number of stages is large, for both the second-order case and the third-order case. In particular, in the second-order case, both the Type-1 and Type-2 permutation filters have the same parameters $\V{\beta}$, and we see that their performance is very close to that of the optimal solution, which is obtained by directly solving \eqref{optimization_problem} relying on the full \textit{a priori} knowledge of the spectral density $f(\lambda)$.

\begin{figure}
\centering
\includegraphics[width=.48\textwidth]{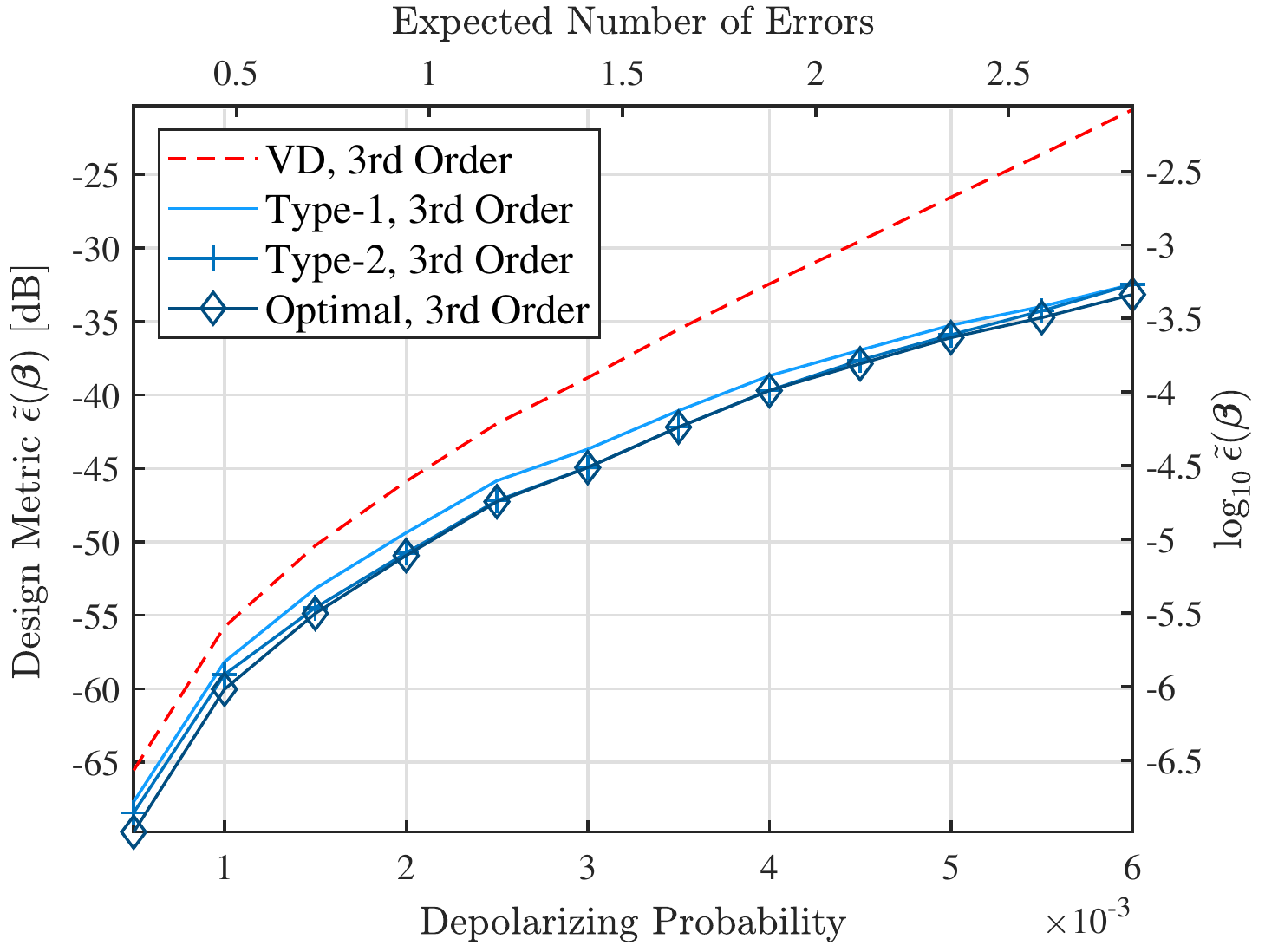}
\caption{The value of the design metric $\tilde{\epsilon}(\V{\beta})$ in \eqref{obj_pareto} for both \ac{vd} and for the proposed methods, as functions of the depolarizing probability. }
\label{fig:abs_error_varying}
\end{figure}

For the third-order case, we see that the Type-2 filter slightly outperforms the Type-1 filter, when the number of stages is relatively small. Intuitively, by adjusting the two zeros of the third-order filters, it is indeed possible to achieve a better error-reduction performance than that of simply placing the zeros at the same point. However, the effect of adjusting the positions of zeros would be less significant when the noise bandwidth is smaller, corresponding to the case where the number of stages is large. Closer scrutiny reveals that the performance of both the Type-1 and Type-2 third-order filters become similar when the number of stages is large, especially when it is larger than $45$. By contrast, the performance of the Type-2 filter is near-optimal when the number of stages is moderate (around 25-40). This trend may prevail, because the Pareto fit becomes more accurate, when the noise bandwidth is moderate.

In Fig.~\ref{fig:error_ub_high}, we consider the case where the two-qubit depolarizing probability is $5\times 10^{-3}$, which is four times that of Fig.~\ref{fig:error_ub_low}. The trends of the curves are similar to those of the lower depolarizing probability scenario. It may now be seen more clearly that the Type-2 permutation filter substantially outperforms its Type-1 counterpart, when the number of stages is small.

Next, in Fig.~\ref{fig:abs_error_varying}, we consider circuits having varying depolarizing probabilities. The number of stages is fixed to $10$. We observe a similar increasing gap between the permutation filters and \ac{vd}. In addition, the Type-2 permutation filter also exhibits better performance for moderate depolarizing probabilities.

\begin{figure}
\centering
\includegraphics[width=.48\textwidth]{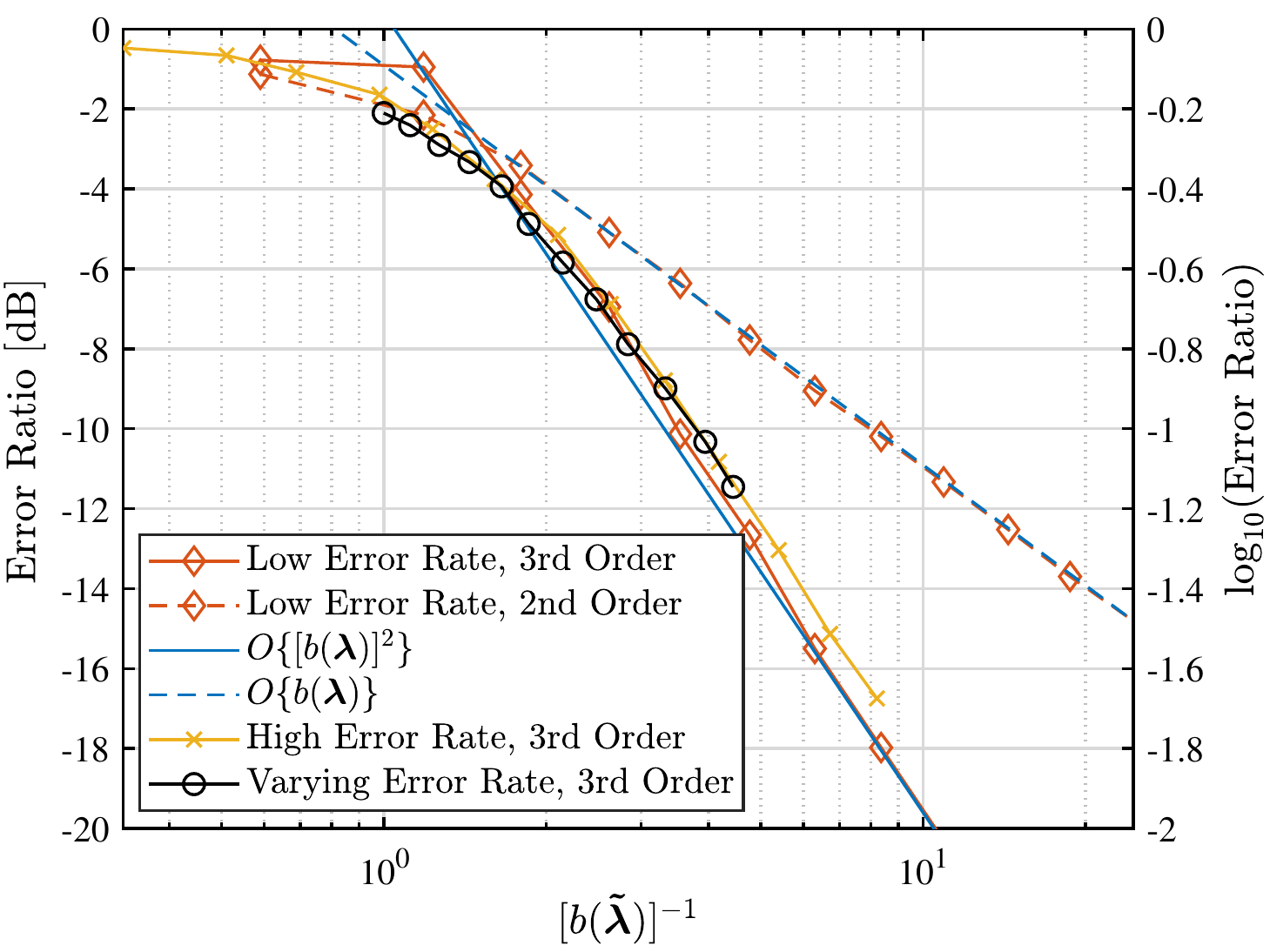}
\caption{The error ratio $R(\V{\beta})$ in \eqref{def_ratio} between Type-1 permutation filters and \ac{vd} vs. the reciprocal of the relative noise bandwidth $b(\tilde{\V{\lambda}})$.}
\label{fig:abs_error_bandwidth}
\end{figure}

In Fig.~\ref{fig:abs_error_bandwidth}, we illustrate the scaling behaviour of the error ratio between the type-1 permutation filters and \ac{vd}, which has been discussed in Section~\ref{sec:performance}. In particular, we plot the error ratios computed using the data presented in Figures \ref{fig:error_ub_low}, \ref{fig:error_ub_high} and \ref{fig:abs_error_varying}. We observe that when the relative noise bandwidth $b(\tilde{\V{\lambda}})$ is small (less than around $0.5$), all error ratios are reduced roughly polynomially with $[b(\tilde{\V{\lambda}})]^{-1}$. Furthermore, the slopes of the curves are almost equal to the asymptotes scaling quadratically and linearly with $b(\tilde{\V{\lambda}})$, respectively for third-order and second-order filters. These observations corroborate Propositions \ref{prop:main} and \ref{prop:pareto}.

\begin{figure}
\centering
\subfloat[][The relative noise bandwidth]{
\centering
\includegraphics[width=.46\textwidth]{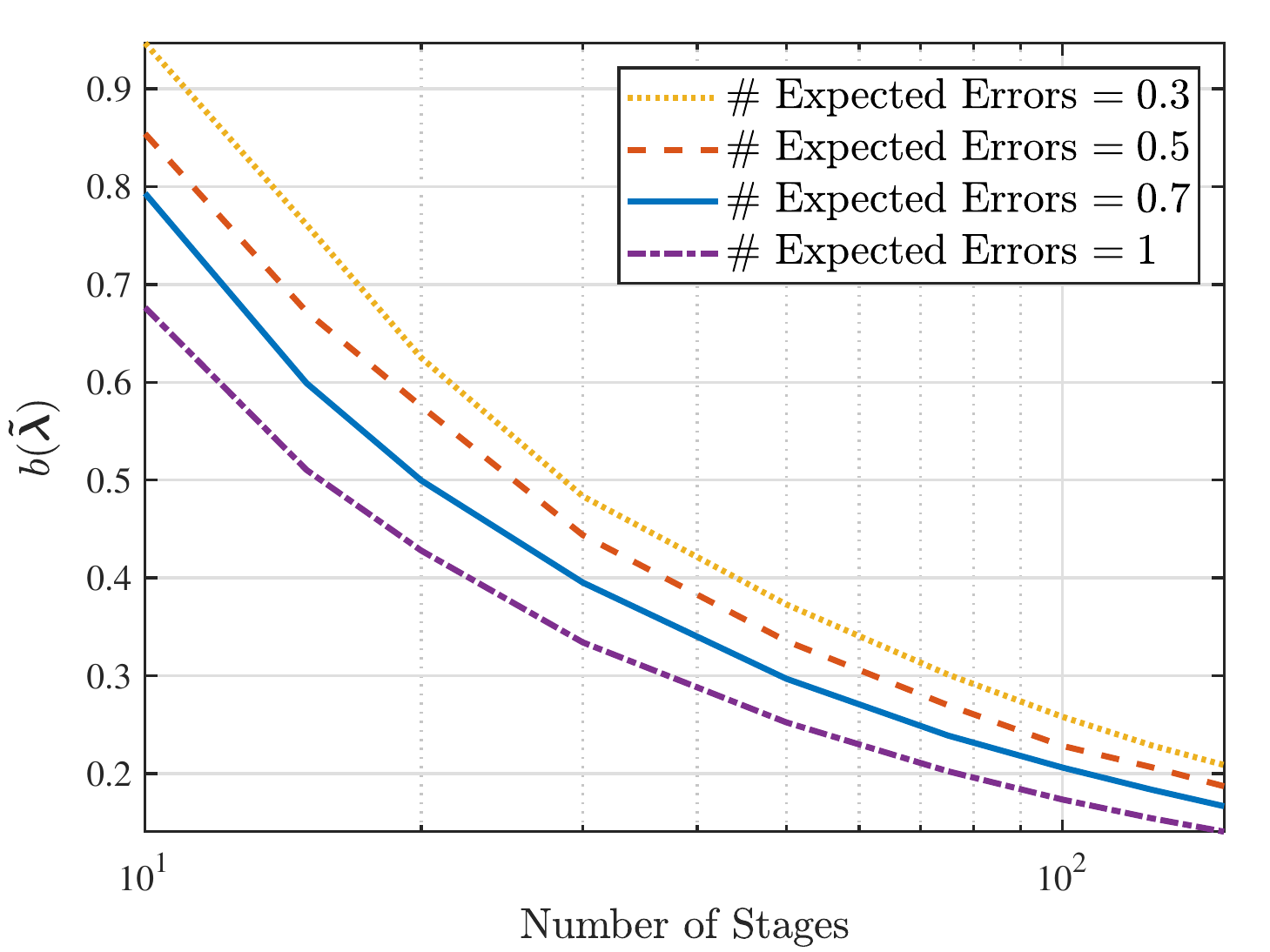}
\label{fig:fixed_expected_error_bandwidth}
}
\\
\subfloat[][The error ratio (The expected number of errors is $0.7$)]{
\centering
\includegraphics[width=.46\textwidth]{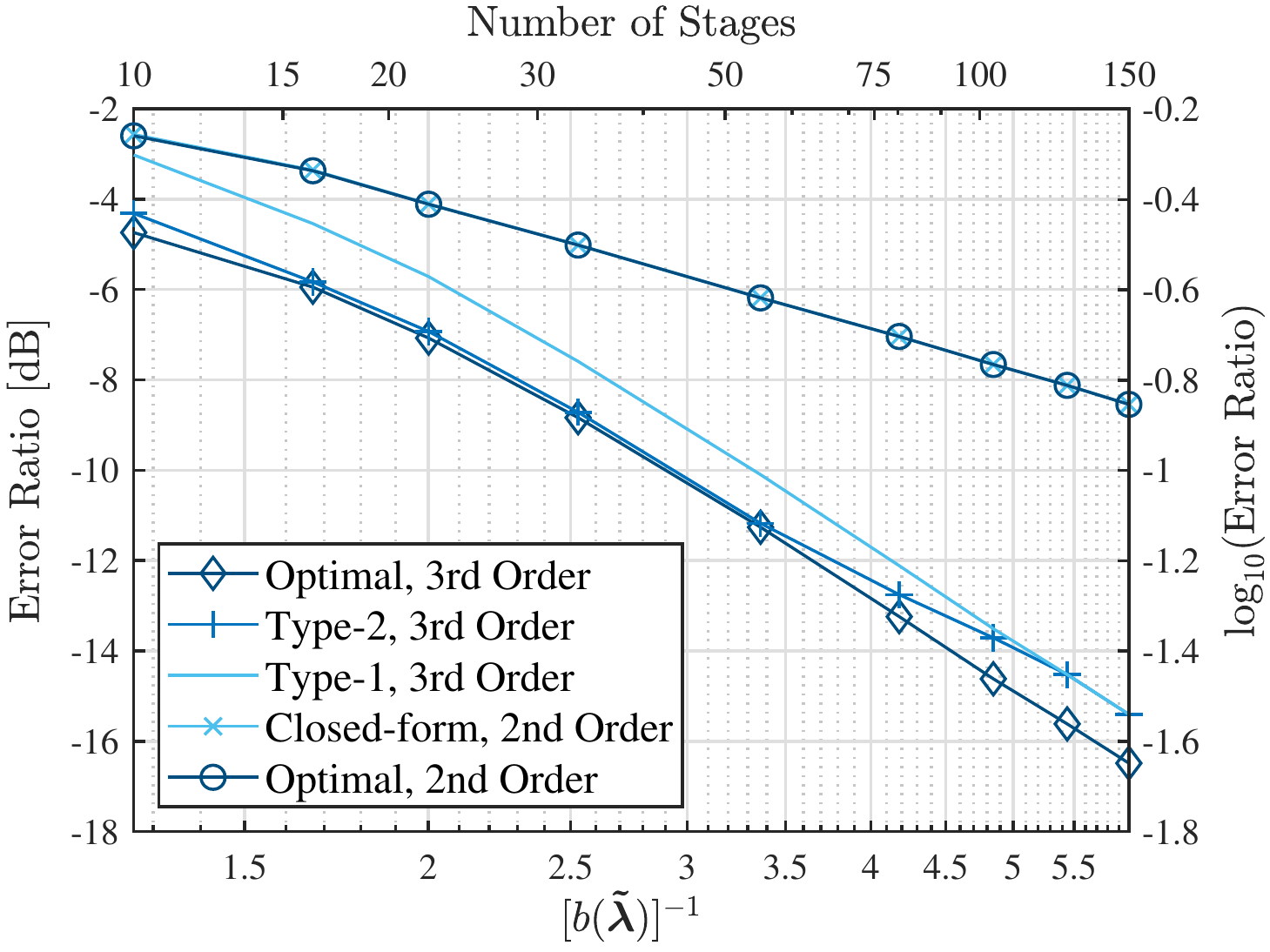}
\label{fig:fixed_expected_error_ratio}
}
\caption{The relative noise bandwidth $b(\tilde{\V{\lambda}})$, and the error ratio $R(\V{\beta})$ in \eqref{def_ratio} between permutation filters and \ac{vd}.}
\label{fig:fixed_expected_error}
\end{figure}

Finally, in Fig.~\ref{fig:fixed_expected_error}, we demonstrate that the commonly used metric of noisiness, namely the expected number of errors, does not determine the relative noise bandwidth on its own, and hence does not solely determine the error ratio between permutation filters and \ac{vd}. To this end, we fixed the expected number of errors, and change both the number of stages as well as the depolarizing probability accordingly. As it can be seen from Fig. \ref{fig:fixed_expected_error_bandwidth}, the relative bandwidth shrinks with the number of stages, even when the expected number of errors is fixed. Similarly, we observe from Fig. \ref{fig:fixed_expected_error} that the error ratio decreases upon reducing the depolarizing probability (or increasing the number of stages).

We may conclude from the discussions in this subsection that the benefit of the permutation filter method is more significant when the circuit is rather noisy, or it is deep but is constituted by gates having relatively small error probabilities.

\subsection{Case Study: \ac{qaoa}-Aided Multi-User Detection}\label{ssec:qaoamud}
In this subsection we demonstrate the performance of permutation filters when applied to a practical variational quantum algorithm, namely the \ac{qaoa}. The parametric state-preparation circuits of \ac{qaoa} are multi-stage circuits having an alternating structure, which take a plus state $\ket{+}^{\otimes N_{\rm q}}$ as the input and produce the following output
\begin{equation}
\ket{\psi}_{\rm out}=e^{-\imath b_{N_{\rm L}}\Sop{H}_{\rm M}}e^{-\imath c_{N_{\rm L}}\Sop{H}_{\rm P}}\dotsc e^{-\imath b_1\Sop{H}_{\rm M}}e^{-\imath c_1\Sop{H}_{\rm P}}\ket{+}^{\otimes N_{\rm q}},
\end{equation}
where $N_{\rm L}$ denotes the number of stages, $\Sop{H}_{\rm M}$ denotes the mixing Hamiltonian defined as $\Sop{H}_{\rm M}:= \sum_{n=1}^{N_{\rm q}} \Sop{X}_i$ ($\Sop{X}_i$ denotes the Pauli-X operator acting on the $i$-th qubit), and $\Sop{H}_{\rm P}$ denotes the phase Hamiltonian that encodes the problem to be solved. The parameters $\V{b}=[b_1,\dotsc,b_{N_{\rm L}}]^{\rm T}$ and $\V{c}=[c_1,\dotsc,c_{N_{\rm L}}]^{\rm T}$ control the dynamic of the algorithm, and are typically determined by an iterative optimization procedure \cite{qaoa}. Since we focus on the performance evaluation for error mitigation methods, here we consider a suboptimal linear scheduling \cite{adiabatic} given by $c_{\ell} = \ell/N_{\rm L}$ and $b_\ell=1-\ell/N_{\rm L}$, instead of optimizing for the parameters.

In particular, we construct the phase Hamiltonian corresponding to the multi-user detection problem \cite{mud} for wireless communication systems.\footnote{For readers not familiar with wireless communication, just note that it is a quadratic unconstrained binary optimization problem.} For an $m\times n$ \ac{mimo} system, the received signal may be modelled as
$$
\V{y}=\M{H}\V{x}+\V{\omega},
$$
where $\V{H}$ denotes the \ac{mimo} channel, $\V{x}$ represents the transmitted signal, and $\V{\omega}$ denotes the noise. For simplicity of the illustration, we assume that the noise is i.i.d. Gaussian on each receiver antenna, and that the modulation scheme is binary phase-shift keying (BPSK), hence $\V{x}\in\{-1,1\}^n$ and $\M{H}\in\mathbb{R}^{m\times n}$. The maximum likelihood estimator of $\V{x}$ can be obtained by solving the following optimization problem
$$
\hat{\V{x}}_{\rm ML} = \mathop{\arg\min}_{\V{x}\in\{-1,1\}^n} \|\V{y}-\M{H}\V{x}\|^2.
$$
The corresponding phase Hamiltonian is thus given by
\begin{equation}
\sum_{k=1}^n [\M{H}^{\rm T}\V{y}]_i \Sop{Z}_i-\sum_{i=1}^{n-1}\sum_{j>i}[\M{H}^{\rm T}\M{H}]_{i,j}\Sop{Z}_i\Sop{Z}_j.
\end{equation}
We consider the following scenario for our numerical simulations: $N_{\rm q}=m=n=10$, the channel $\M{H}$ has i.i.d Gaussian entries with zero mean and a variance of $1/m=0.1$, and the signal-to-noise ratio is $13$dB, implying that $[\V{\omega}]_i\sim\Sop{N}(0,0.05)$.

\begin{figure}
\centering
\subfloat[][The computational error]{
\centering
\includegraphics[width=.46\textwidth]{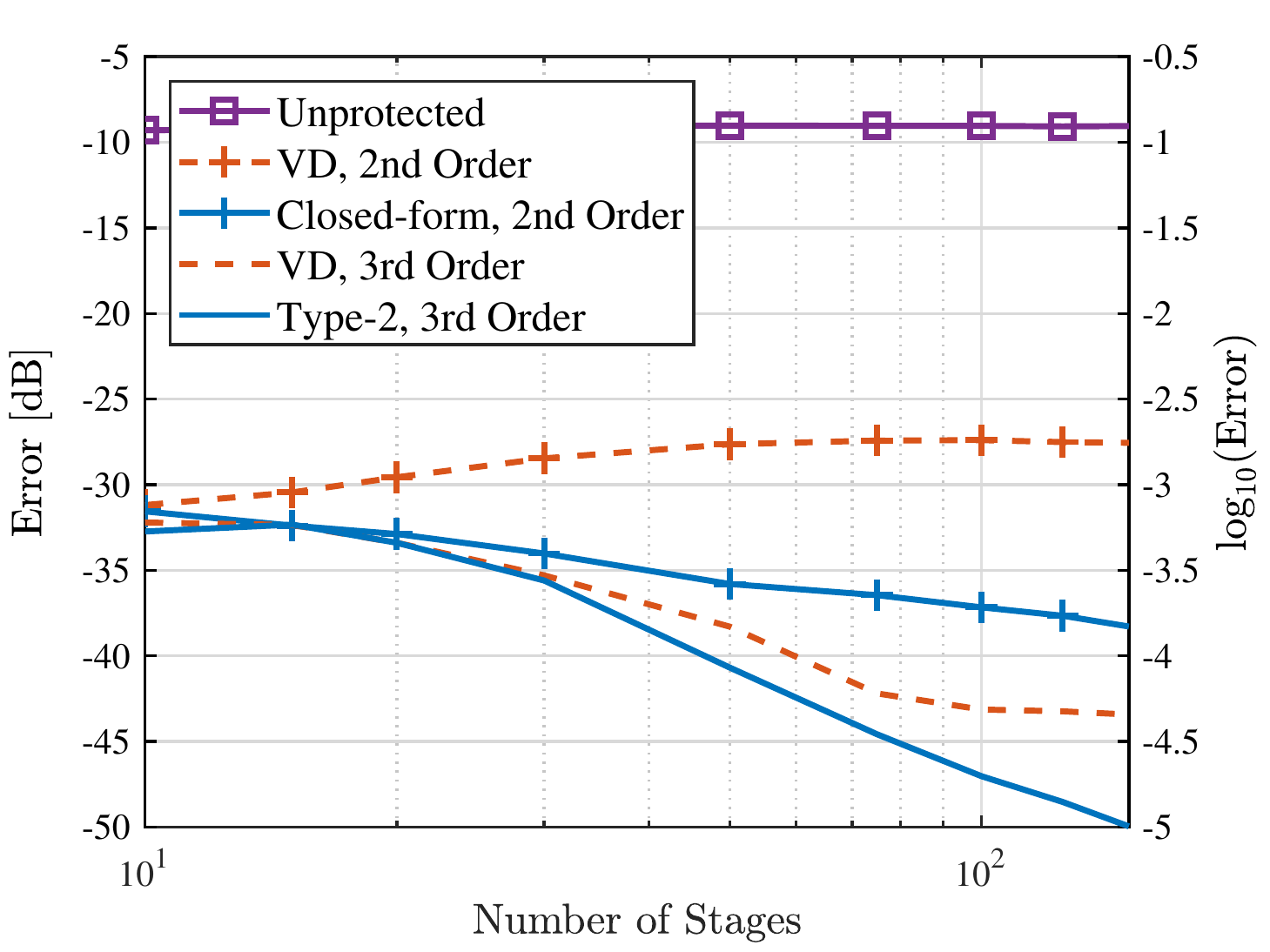}
\label{fig:fixed_expected_error_qaoamud}
}
\\
\subfloat[][The sampling overhead factor]{
\centering
\includegraphics[width=.46\textwidth]{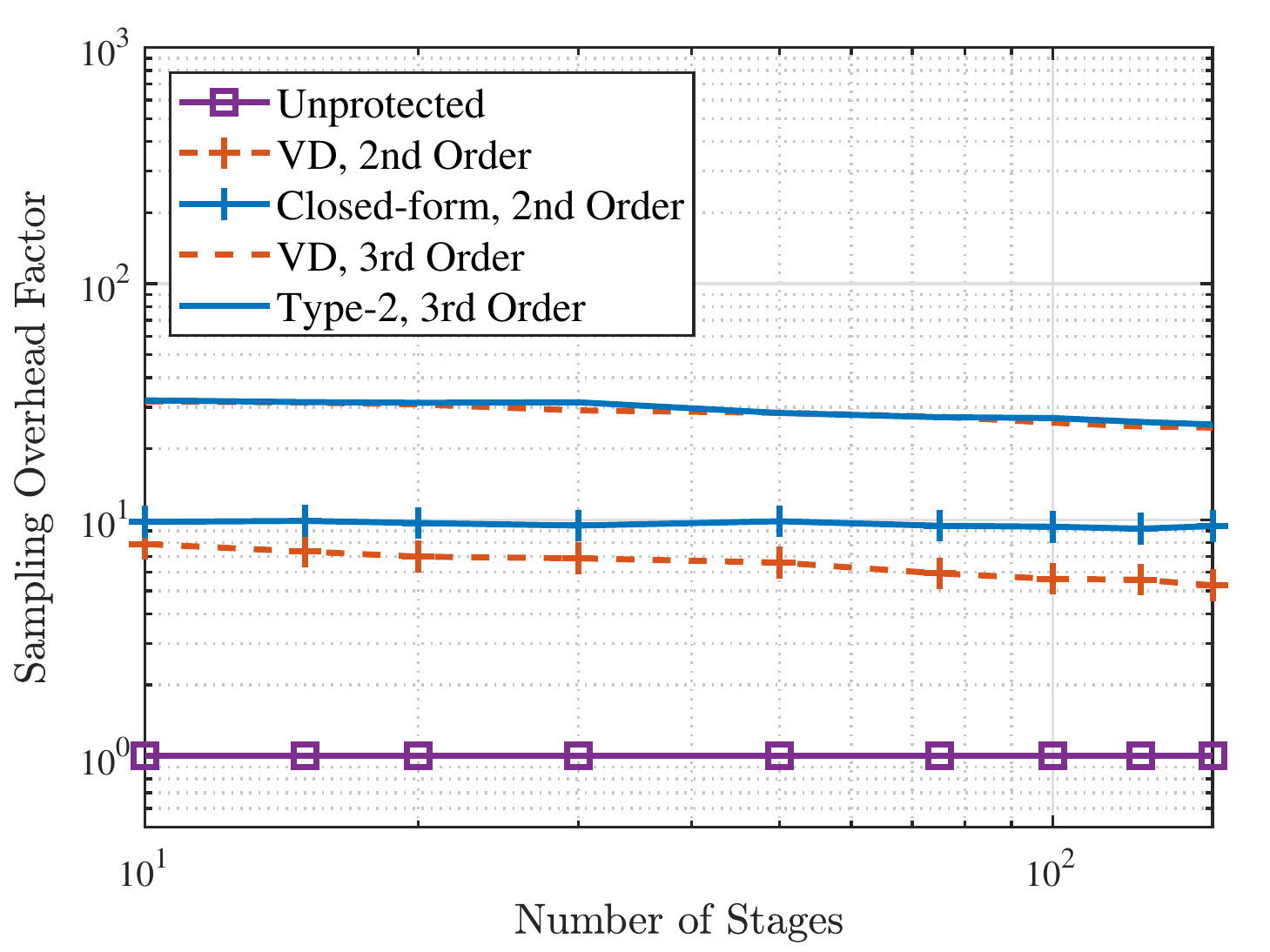}
\label{fig:fixed_expected_error_qaoamud_sof}
}
\caption{The computational error and the sampling overhead factor of permutation filters applied to \ac{qaoa}-aided multi-user detection vs. the number of stages, where the expected number of errors is fixed at $0.7$.}
\label{fig:fixed_expected_error_qaoamuds}
\end{figure}

\begin{figure}
\centering
\subfloat[][The computational error]{
\centering
\includegraphics[width=.46\textwidth]{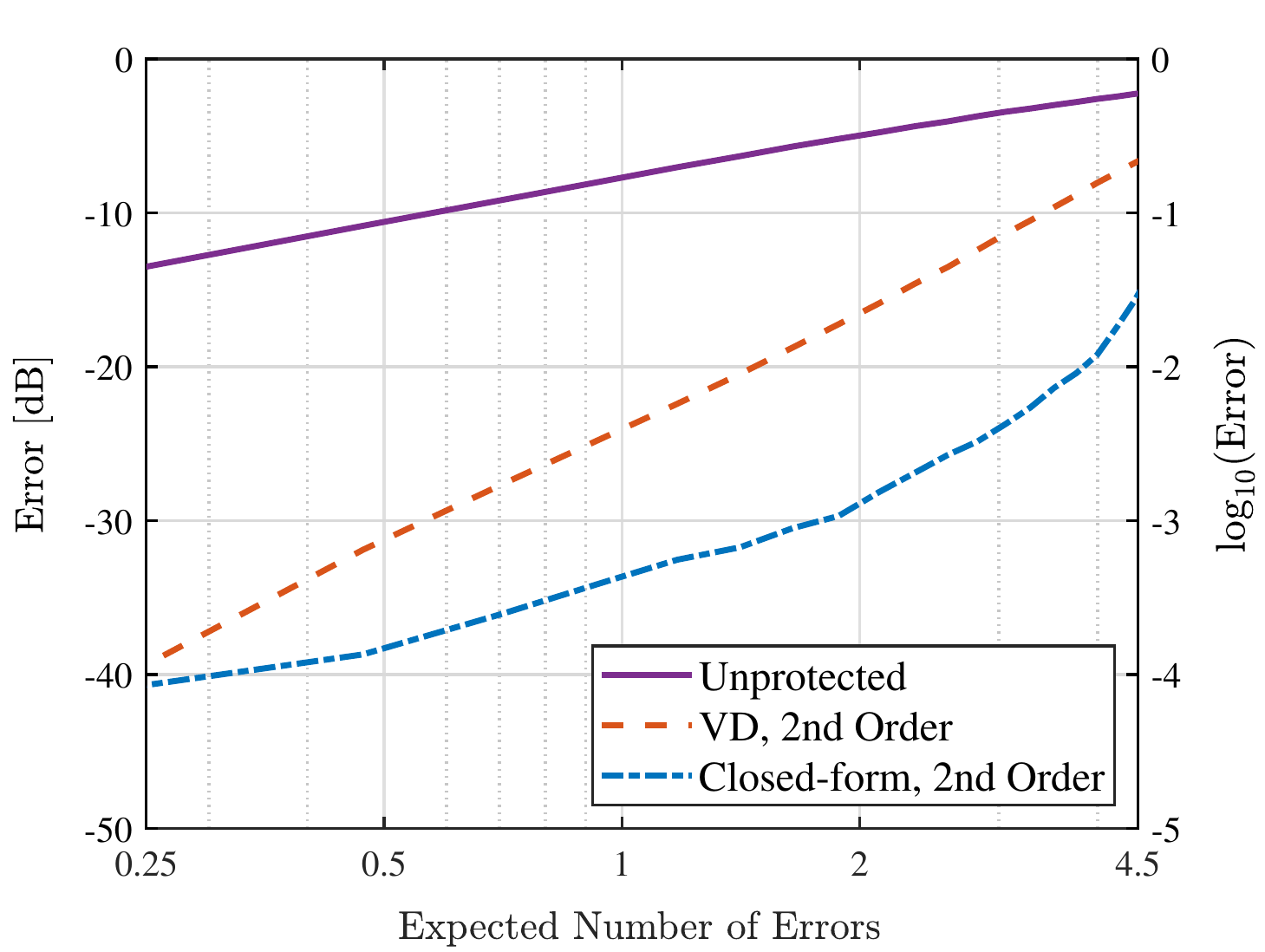}
\label{fig:errrate_qaoamud}
}
\\
\subfloat[][The sampling overhead factor]{
\centering
\includegraphics[width=.46\textwidth]{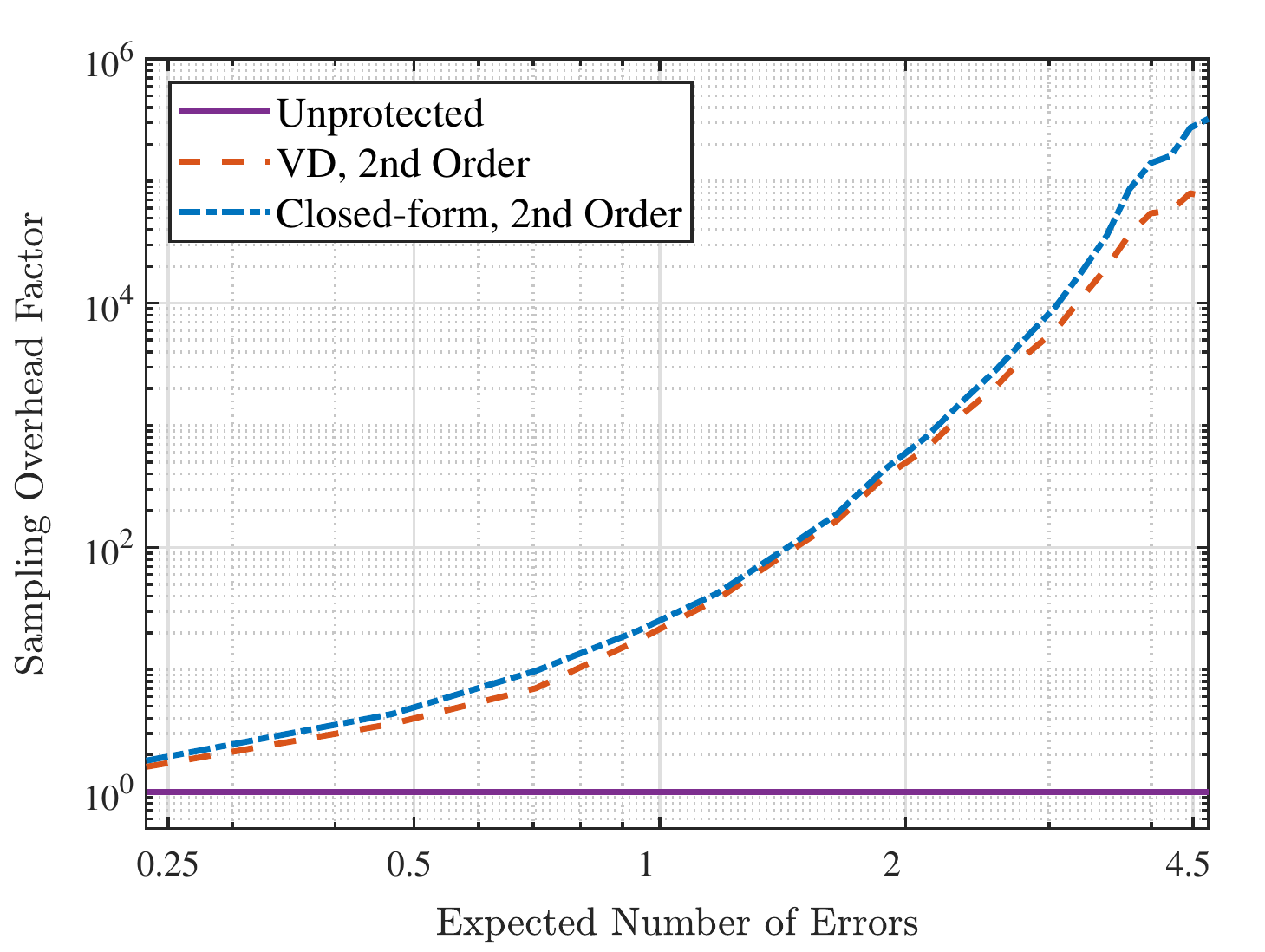}
\label{fig:errrate_qaoamud_sof}
}
\caption{The computational error and the sampling overhead factor of permutation filters applied to  \ac{qaoa}-aided multi-user detection vs. the expected number of errors, where the number of stages is fixed at $50$.}
\label{fig:errrate_qaoamuds}
\end{figure}

We first fix the expected number of errors at $0.7$ and investigate the dependency of the computational error (the absolute difference between the error-free result and the result computed relying on noisy circuits based on the entire Hamiltonian) on the number of stages. As it may be seen from Fig.~ \ref{fig:fixed_expected_error_qaoamud}, the permutation filters are more beneficial when the circuit is deep, as have been discussed in Section \ref{ssec:abs_error}. We may also observe from Fig.~\ref{fig:fixed_expected_error_qaoamud_sof} that the sampling overhead is nearly constant with the number of stages, suggesting that the expected number of errors might be the principal determining factor of the overhead.

Next, we present the relationship between the computational error and the expected number of error, with a fixed number of stages $N_{\rm L}=50$, in Fig. \ref{fig:errrate_qaoamud}. It is seen from the figure that the permutation filter improves the error mitigation performance significantly when the expected number of errors is large. However, it should also be noted that the sampling overhead increases dramatically when the expected number of errors is larger than $1$, as shown in Fig.~\ref{fig:errrate_qaoamud_sof}. Extra care should be taken for this issue, since a high sampling overhead may render the error mitigation method unfavorable in practice.

\section{Conclusions}\label{sec:conclusions}
In this treatise, we have proposed a general framework for designing \ac{fir}-like permutation filters for mitigating the computational errors of variational quantum algorithms. In particular, the filter design problem is an invex problem, hence the algorithm is guaranteed to converge to the global optimum. For narrowband noise scenarios, we have also shown a polynomial error reduction compared to \ac{vd}. This implies that permutation filters improve the error-reduction performance more substantially for quantum circuits having large depth or higher gate error rate.

The performance metric we used for filter design is an upper bound of the error magnitude across all unitary observables. A possible future research direction is to find other metrics better suited to specific classes of practical observables.

\appendices
\section{Proof of Proposition \ref{prop:global_minimum}}\label{sec:proof_global_minimum}
\begin{IEEEproof}
Consider the transform from $\V{\beta}$ to $\V{\alpha}$, which helps us to reformulate \eqref{optimization_problem} (where the cost function is approximated as in \eqref{approx_beta}) in the form of:
\begin{equation}\label{optimization_alpha}
\begin{aligned}
\min_{\V{\alpha}} &~~\xi(\V{\alpha}),\\
{\rm s.t.}&~~[\V{\alpha}]_1=1,
\end{aligned}
\end{equation}
where
\begin{equation}
\xi(\V{\alpha}):=\tilde{\epsilon}[\V{\varphi}(\V{\alpha})]=\int _{\lambda_{\rm m}}^1 f(\lambda)\sqrt{\V{\alpha}^{\rm T}\M{A}(\lambda)\V{\alpha}}~{\rm d}\lambda,
\end{equation}
$\V{\varphi}(\cdot)$ is the mapping from $\V{\beta}$ to $\V{\alpha}$, $\M{A}(\lambda)$ is defined by $\M{A}(\lambda):=\V{a}(\lambda)[\V{a}(\lambda)]^{\rm T}$, and $\V{a}(\lambda):=[\lambda^N~\lambda^{N-1}~\dotsc~\lambda]^{\rm T}$. Note that the term $\sqrt{\V{\alpha}^{\rm T}\M{A}(\lambda)\V{\alpha}}$ is actually the Mahalanobis norm \cite{maha_norm} of $\V{\alpha}$ with respect to a positive semi-definite symmetric matrix $\M{A}(\lambda)$, hence it is a convex function of $\V{\alpha}$. Thus the objective function itself is also convex with respect to $\V{\alpha}$, since the integration (weighted by a non-negative function $f(\lambda)$) preserves convexity.

Next, we observe that $\V{\varphi}(\cdot)$ can be computed via \eqref{def_alpha}, and its inverse may be obtained using the factorization of polynomials \cite{poly_factoring}. Since $\V{\beta}$ satisfies the ordering \eqref{order_zeros}, when $\V{\alpha}$ is further constrained to be the coefficients of polynomials having only non-negative real-valued roots, it is clear that $\V{\varphi}(\cdot)$ is a bijection, and hence the Jacobian $\M{J}_{\V{\beta}}$ that is given by
$$
\M{J}_{\V{\beta}}=\left[\frac{\partial \V{\varphi}(\V{\beta})}{\partial \beta_1}~\frac{\partial \V{\varphi}(\V{\beta})}{\partial \beta_2}~\dotsc ~\frac{\partial \V{\varphi}(\V{\beta})}{\partial \beta_{N-1}}\right]^{\rm T},
$$
is invertible for every $\V{\beta}\in\Set{B}$. This implies that $\V{\varphi}(\cdot)$ is a diffeomorphism from $\V{\beta}$ to $\V{\alpha}$, and hence $\tilde{\epsilon}(\V{\beta})$ is an invex function of $\V{\beta}$ \cite{invex1,invex2,invex_ieee}. To elaborate further, we see that
$$
\left.\frac{\partial \tilde{\epsilon}(\V{\beta})}{\partial \V{\beta}}\right|_{\V{\beta}_0}=\M{J}_{\V{\beta}_0}^{-1}\left.\frac{\partial \xi(\V{\alpha})}{\partial \V{\alpha}}\right|_{\V{\varphi}(\V{\beta}_0)}=\V{0}\Leftrightarrow \left.\frac{\partial \xi(\V{\alpha})}{\partial \V{\alpha}}\right|_{\V{\varphi}(\V{\beta}_0)}=\V{0}
$$
holds for $\V{\beta}_0\in\Set{B}$, implying that $\V{\beta}_0\in\Set{B}$ is a stationary point of $\tilde{\epsilon}(\V{\beta})$ if and only if $\V{\varphi}(\V{\beta}_0)$ is also a stationary point of $\xi(\V{\alpha})$, which in turn is one of the global minima of $\xi(\V{\alpha})$.

Our remaining task is to show that $\xi(\V{\alpha})$ attains its global minimum when $\V{\beta}=\V{\varphi}^{-1}(\V{\alpha})$ belongs to the feasible region $\Set{B}$. This may be proved using the method of contradiction. Assume by contrast that the minimum of $\xi(\V{\alpha})$ is attained at $\V{\alpha}_0\notin\Set{B}$. Then the polynomial $\V{\alpha}_0^{\rm T}\V{a}(\lambda)$ has either real negative roots or complex roots. For the former case, it is plausible that $|\V{\alpha}_0^{\rm T}\V{a}(\lambda)|>\lambda^N$ for all $\lambda>0$, hence $\V{\alpha}_0$ is not the optimum. For the latter case, we specifically consider a pair of conjugate complex roots $x\pm i y$. It is clear that
$$
\begin{aligned}
|(\lambda-x-iy)(\lambda-x+iy)|&=\lambda^2 -2x \lambda +\sqrt{x^2+y^2}\\
&\ge \lambda^2 -2x \lambda + x^2 =(\lambda-x)^2,
\end{aligned}
$$
implying that the cost function value can be reduced by replacing the complex roots with real roots. Hence the proof is completed.
\end{IEEEproof}

\section{Notes on the Spectral Response of \\Permutation Filters}\label{sec:notes_spectral_response}
Let us consider a third-order permutation filter as an example, which has the following spectral response:
\begin{equation}
h_{\V{\beta}}(\lambda) = \lambda(\lambda-\beta_1)(\lambda-\beta_2),
\end{equation}
where $\beta_1 \le \beta_2$. By taking the limit $\lambda\rightarrow \infty$, we see that $h_{\V{\beta}}(\lambda)\sim \lambda^3$, implying that the spectral response can be well approximated by $\lambda^3$ when $\lambda \gg \beta_2$. Since the cubic function $\lambda^3$ satisfies $\lambda_1^3=10^3\cdot\lambda_2^3$ when $\lambda_1=10\lambda_2$, we say that it ``has a slope of $30$dB per decade'' (note that $10$dB corresponds to $10\log_{10}(10)=10$ times). Here, the ``slope'' refers to that of the spectral response curve on a log-log scale, which appears to be linear for power functions. Furthermore, if $\beta_1$ and $\beta_2$ is well separated, we see that $h_{\V{\beta}}(\lambda)\sim \lambda^2$ when $\beta_1\ll \lambda\ll \beta_2$, and hence ``has a slope of $20$dB per decade''. In general, when the eigenvalue $\lambda$ is in the region $\beta_n \ll \lambda \ll \beta_{n+1}$, we see that the slope is (approximately) $10(n+1)$~dB per decade. Since the first zero is $\beta_0=0$, we may conclude that each zero $\beta_i\ll \lambda_0 $ contributes $10$dB/decade to the slope at the point $\lambda = \lambda_0$.

\section{Proof of Proposition \ref{prop:pareto}}\label{sec:proof_pareto}
\begin{IEEEproof}
The term $\tilde{\epsilon}(\V{0})$ may be viewed as the $N$-th moment of the Pareto distribution. Upon denoting the shape parameter and the minimum value of the Pareto distribution as $k$ and $\lambda_{\rm m}$, we have
\begin{equation}
\tilde{\epsilon}(\V{0})=\frac{k}{k-N}\cdot\lambda_{\rm m}^N.
\end{equation}
From \eqref{obj_pareto} we obtain
\begin{equation}\label{pareto_scaling}
\begin{aligned}
R(\V{\beta})&=\frac{\tilde{\epsilon}(\V{\beta})}{\tilde{\epsilon}(\V{0})}\\
&=\frac{k-N}{k\lambda_{\rm m}^N}\sum_{i=0}^{N-1}(-1)^i\int_{\beta_{N-i-1}}^{\beta_{N-i}}G_{\V{\beta}}(\lambda) {\rm d}\lambda.
\end{aligned}
\end{equation}
Note that for Type-1 permutation filters, we have $\V{\beta}=\frac{k\lambda_{\rm m}}{k-1}\V{1}$. Hence \eqref{pareto_scaling} can be bounded as
\begin{equation}\label{upperbound_ratio}
\begin{aligned}
R(\V{\beta})&=\frac{k-N}{k\lambda_{\rm m}^N}\left(\int_{\frac{k\lambda_{\rm m}}{k-1}}^\infty |G_{\V{\beta}}(\lambda)|{\rm d}\lambda + \int_{\lambda_{\rm m}}^{\frac{k\lambda_{\rm m}}{k-1}} |G_{\V{\beta}}(\lambda)|{\rm d}\lambda\right) \\
&=\left|G(\lambda_{\rm m})-G\left(\frac{k\lambda_{\rm m}}{k-1}\right)\right|+\left|G\left(\frac{k\lambda_{\rm m}}{k-1}\right)\right|\\
&\le 2\left|G\left(\frac{k\lambda_{\rm m}}{k-1}\right)\right|,
\end{aligned}
\end{equation}
where for simplicity of notations we have defined $G(\lambda)=\frac{k-N}{\lambda_{\rm m}^{N-k}}\tilde{G}_{\V{\alpha}}(\lambda)$. The last line of \eqref{upperbound_ratio} comes from the fact that
$$
\int_{\lambda_{\rm m}}^{\frac{k\lambda_{\rm m}}{k-1}} |G_{\V{\beta}}(\lambda)|{\rm d}\lambda\ge 0.
$$
Furthermore, from \eqref{def_alpha} we have
\begin{equation}
\begin{aligned}
\alpha_i &=\tbinom{N-1}{i-1} \left(-\frac{k\lambda_{\rm m}}{k-1}\right)^{i-1}.
\end{aligned}
\end{equation}
Thus we obtain
\begin{equation}
\begin{aligned}
G(\lambda)&=\sum_{n=1}^N \alpha_{N-n+1}\frac{k-N}{n-k}\cdot \frac{\lambda^{n-k}}{\lambda_{\rm m}^{N-k}} \\
&=(k-N)\sum_{n=1}^N\frac{ \tbinom{N-1}{N-n}}{n-k} \left(\frac{-k}{k-1}\right)^{N-n}\left(\frac{\lambda}{\lambda_{\rm m}}\right)^{n-k}.
\end{aligned}
\end{equation}
This implies that
\begin{equation}\label{pre_eta}
G\left(\frac{k\lambda_{\rm m}}{k-1}\right)=\frac{k-N}{\left(\frac{k}{k-1}\right)^{k-N}}\sum_{n=1}^N\frac{ \tbinom{N-1}{n-1}}{n-k} (-1)^{N-n}.
\end{equation}

Next, we denote
\begin{equation}
\begin{aligned}
\sum_{n=1}^N(-1)^{N-n}\tbinom{N-1}{n-1}\eta(n,k)=\V{a}_{N-1}^{\rm T}\V{\eta},
\end{aligned}
\end{equation}
where $[\V{a}_{N-1}]_i=(-1)^{N-i}\tbinom{N-1}{i-1}$, $[\V{\eta}]_i=\eta(i,k)$, and $\eta(n,k)$ denotes an arbitrary function of $n$ and $k$. Furthermore, we have
\begin{equation}
\V{a}_{N-1}^{\rm T}\V{\eta}=\V{1}^{\rm T}\M{A}_{N-1}\V{\eta},
\end{equation}
where $\M{A}_{N-1}$ is defined recursively by
\begin{equation}
\M{A}_n=\left[
              \begin{array}{cc}
                \M{A}_{n-1} & \V{0}_{2^{n-2}\times 1} \\
                 \V{0}_{2^{n-2}\times 1} & -\M{A}_{n-1} \\
              \end{array}
            \right],
\end{equation}
and $\M{A}_1:=[1~-1]$. Thus we have the following recursion
$$
\V{1}^{\rm T}\M{A}_n\V{x}=\V{1}^{\rm T}\M{A}_{n-1}\left([\V{x}]_{1:L-1}-[\V{x}]_{2:L}\right)
$$
for $\V{x}\in\mathbb{R}^L$. From \eqref{pre_eta} we may now write $\V{\eta}$ explicitly as
\begin{equation}
\V{\eta}=\left[\frac{1}{1-k}~\frac{1}{2-k}~\dotsc~\frac{1}{N-k}\right]^{\rm T}.
\end{equation}
When $N=2$, we have
$$
\begin{aligned}
\V{1}^{\rm T}\M{A}_1\V{\eta} &= \frac{1}{1-k}-\frac{1}{2-k} \\
&= \frac{\Gamma(-k)}{\Gamma(1-k)}-\frac{\Gamma(1-k)}{\Gamma(2-k)},
\end{aligned}
$$
where $\Gamma(\cdot)$ denotes the Gamma function \cite{table_integral}. Note that
\begin{equation}
\begin{aligned}
\frac{\Gamma(-k)}{\Gamma(m-k)}-\frac{\Gamma(1-k)}{\Gamma(m-k+1)}=\frac{m\Gamma(-k)}{\Gamma(m+1-k)}.
\end{aligned}
\end{equation}
For $N=3$ we obtain
$$
\begin{aligned}
\V{1}^{\rm T}\M{A}_2\V{\eta} &= \frac{\Gamma(-k)}{\Gamma(2-k)}-\frac{\Gamma(1-k)}{\Gamma(2-k+1)} =\frac{2\Gamma(-k)}{\Gamma(3-k)},
\end{aligned}
$$
and hence in general we have
$$
\begin{aligned}
\V{1}^{\rm T}\M{A}_{N-1}\V{\eta}&=\frac{(N-1)!\Gamma(-k)}{\Gamma(N-k)}\\
&=(-1)^N(N-1)!\cdot \frac{\Gamma(k-N-1)}{\Gamma(k)}.
\end{aligned}
$$
This implies that
\begin{equation}\label{scaling1}
\begin{aligned}
\left|G\left(\frac{k\lambda_{\rm m}}{k-1}\right)\right|&=\frac{(k-N)(N-1)!}{\left(\frac{k}{k-1}\right)^{k-N}}\cdot \frac{\Gamma(k-N-1)}{\Gamma(k)} \\
&= \frac{\left(1+\frac{1}{k-1}\right)^{N-k}(N-1)!}{\prod_{n=1}^{N-1}(k-n)},
\end{aligned}
\end{equation}
as a function of $k$.

Finally, since we have assumed that the spectral density obeys a Pareto distribution, we may compute the relative noise bandwidth explicitly as follows:
\begin{equation}\label{scaling2}
\begin{aligned}
b(\tilde{\V{\lambda}})&=\sqrt{\frac{k}{(k-1)^2(k-2)}}\\
&\ge (k-1)^{-1}.
\end{aligned}
\end{equation}
Combining \eqref{scaling1} and \eqref{scaling2}, we obtain the desired scaling law in \eqref{desired_pareto}.
\end{IEEEproof}

\section{Proof of Proposition \ref{prop:concentration}}\label{sec:proof_concentration}
\begin{IEEEproof}
To simplify the discussion, we will use the Pauli basis. Under the Pauli basis, a quantum channel $\Sop{C}$ may be represented in a matrix form as
\begin{equation}
[\M{C}]_{i,j} = \frac{1}{2^{N_{\rm q}}}\tr{\Sop{S}_i\Sop{C}(\Sop{S}_j)},
\end{equation}
where $\Sop{S}_i$ denotes the $i$-th Pauli string acting upon $N_{\rm q}$ qubits. Correspondingly, a quantum state $\rho$ may be represented as a vector $[\V{x}_\rho]_i = \frac{1}{\sqrt{2^{N_{\rm q}}}}\tr{\Sop{S}_i\rho}$. Since the Pauli operators are unitary and mutually orthogonal, both the transform from the conventional computation basis to the Pauli basis, as well as the inverse transform, are also unitary. This implies that
\begin{equation}\label{unitary_invariance}
\|\V{\lambda}_{\rho}\|_2=\|\V{x}_\rho\|_{\rm F}=\|\V{x}_\rho\|_2,
\end{equation}
due to the unitary invariance of the Frobenius norm \cite{matrix_analysis}, where $\V{\lambda}_\rho$ denotes the vector containing all eigenvalues of $\rho$ sorted in descending order. Without loss of generality, we assume that the first Pauli operator is the identity operator $\Sop{I}^{\otimes N_{\rm q}}$. In light of this, we have $\V{x}_{\rho}=[2^{-N_{\rm q}/2}~\tilde{\V{x}}_{\rho}^{\rm T}]^{\rm T}$, since all quantum states satisfy $\tr{\rho}=1$.

We say that ``a layer of gates'' is activated if each qubit has been act upon by at least one gate. From our assumption we see that the circuit consists of at least $L$ layers. After the $l$-th layer, the output state $\V{x}_{\rho_l}$ may be expressed as
\begin{equation}
\V{x}_{\rho_l} = \widetilde{\M{G}}_l\V{x}_{\rho_{l-1}}=\M{C}_l\M{G}_l\V{x}_{\rho_{l-1}},
\end{equation}
where $\M{G}_l$ denotes the ideal noiseless operation corresponding to the $l$-th layer, and $\M{C}_l$ denotes the associated quantum channel characterizing the noise. A perfect layer of gates $\M{G}_i$, and the corresponding Pauli channel $\M{C}_i$, can be expressed as
\begin{equation}
\M{G}_i = \left[
            \begin{array}{cc}
              1 & \V{0}^{\rm T} \\
              \V{0} & \M{U}_i \\
            \end{array}
          \right],~~\M{C}_i = \left[
            \begin{array}{cc}
              1 & \V{0}^{\rm T} \\
              \V{0} & \M{D}_i \\
            \end{array}
          \right],
\end{equation}
respectively, where $\M{U}_i\in \mathbb{R}^{(4^{N_{\rm q}}-1)\times (4^{N_{\rm q}}-1)}$ is a unitary matrix, and $\M{D}_i$ is a diagonal matrix, whose diagonal entries take values in the interval $[0,1]$. We now see that the maximum singular value of $\widetilde{\M{G}}_l$ is $1$, while its second largest singular value $\sigma_2(\widetilde{\M{G}}_l)$ is given by
\begin{equation}
\sigma_2(\widetilde{\M{G}}_l) =\|\M{D}_l\|_2.
\end{equation}

Since the probability of each single-qubit Pauli error is at least $\epsilon_{\rm l}$, we see that for a single-qubit channel $\Sop{C}$ characterized by the error probabilities of $p_{\rm X}$, $p_{\rm Y}$ and $p_{\rm Z}$ corresponding to the X, Y and Z errors, respectively, the following holds:

\begin{equation}\label{pauli_channel_ptm}
\begin{aligned}
\M{C} &= \diag{\widetilde{\M{H}} [1-p_{\rm X}-p_{\rm Y}-p_{\rm Z}~~p_{\rm X}~~p_{\rm Y}~~p_{\rm Z}]^{\rm T}} \\
&=\M{I}-2\diag{[p_{\rm X}+p_{\rm Z}~~p_{\rm Y}+p_{\rm Z}~~p_{\rm X}+p_{\rm Y}]} \\
& \preceq (1-4\epsilon_{\rm l})\M{I},
\end{aligned}
\end{equation}
where $\widetilde{\M{H}}$ denotes the inverse Hadamard transform over $N_{\rm q}$ qubits. Therefore, we obtain
\begin{equation}
\begin{aligned}
\|\tilde{\V{x}}_{\rho_{L}}\|_2&\le \sigma_2\left(\prod_{l=1}^{L} \widetilde{\M{G}}_{L-l+1}\right) \\
&\le \prod_{l=1}^{L} \|\M{D}_l\|_2 \\
&\le (1-4\epsilon_{\rm l})^{L} \\
&\le \exp\left(-4\epsilon_{\rm l}L\right),
\end{aligned}
\end{equation}
where the last line follows from the fact that $\ln(1-x) \le -x$ holds for all $x>0$. This implies that
\begin{equation}
\|\V{x}_{\rho_L}-[2^{N_{\rm q}/2}~\V{0}^{\rm T}]^{\rm T}\|_2\le \exp\left(-4\epsilon_{\rm l}L\right).
\end{equation}

Note that $[2^{-N_{\rm q}/2}~\V{0}^{\rm T}]^{\rm T}$ corresponds to the completely mixed state $2^{-N_{\rm q}}\M{I}$, hence from \eqref{unitary_invariance} we have
\begin{equation}
\begin{aligned}
\|\V{\lambda}_{\rho_{L}}-2^{-N_{\rm q}}\V{1}\|_2&=\|\rho_L-2^{-N_{\rm q}}\M{I}\|_{\rm F} \\
&\le \exp\left(-4\epsilon_{\rm l}L\right).
\end{aligned}
\end{equation}
The relative noise bandwidth is given by
\begin{equation}\label{fbw}
b(\tilde{\V{\lambda}}) = \mu^{-1}(2^{N_{\rm q}}-1)^{-\frac{1}{2}}\|\tilde{\V{\lambda}}-\mu\V{1}\|_2,
\end{equation}
where $\mu=\frac{1-[\V{\lambda}_{\rho_{L}}]_1}{2^{N_{\rm q}}-1}$ and $\tilde{\V{\lambda}} = [\V{\lambda}_{\rho_{L}}]_{2:2^{N_{\rm q}}}$. The term $\|\tilde{\V{\lambda}}-(2^{N_{\rm q}}-1)^{-1}\V{1}|_2$ can be bounded by
\begin{equation}\label{up_bw}
\begin{aligned}
\|\tilde{\V{\lambda}}-\mu\V{1}\|_2&\le \|\tilde{\V{\lambda}}-2^{-N_{\rm q}}\V{1}\|_2+\frac{\left\|[\V{\lambda}_{\rho_{L}}]_1-2^{-N_{\rm q}}\V{1}\right\|_2}{2^{N_{\rm q}}-1}\\
&\le \left(1+(2^{N_{\rm q}}-1)^{-1/2}\right)e^{-4\epsilon_{\rm l}L}.
\end{aligned}
\end{equation}
In addition, we have
\begin{equation}\label{lb_mu}
\begin{aligned}
\mu&= \frac{1-2^{-N_{\rm q}}-|[\V{\lambda}_{\rho_{L}}]_1-2^{-N_{\rm q}}|}{2^{N_{\rm q}}-1}\\
&\ge\frac{1-2^{-N_{\rm q}}-e^{-4\epsilon_{\rm l}L}}{2^{N_{\rm q}}-1}.
\end{aligned}
\end{equation}
Substituting \eqref{up_bw} and \eqref{lb_mu} into \eqref{fbw}, we obtain \eqref{exponential_concentration}. Thus the proof is completed.
\end{IEEEproof}

\bibliographystyle{ieeetran}
\bibliography{IEEEabrv,QEM}
\end{document}